\documentclass[a4paper,fleqn]{cas-dc}

\usepackage[sort&compress,numbers]{natbib}
\usepackage[utf8]{inputenc}
\usepackage{textcomp}
\DeclareUnicodeCharacter{22EF}{\dots}
\DeclareUnicodeCharacter{0302}{\^}

\usepackage[version=4]{mhchem} 
\usepackage{siunitx} 

\usepackage{soul}

\def\tsc#1{\csdef{#1}{\textsc{\lowercase{#1}}\xspace}}
\tsc{WGM}
\tsc{QE}
\tsc{EP}
\tsc{PMS}
\tsc{BEC}
\tsc{DE}

\begin{document}
\let\WriteBookmarks\relax
\def\floatpagepagefraction{1}
\def\textpagefraction{.001}
\shorttitle{Petal-Graphyne: A Novel 2D Carbon Allotrope}
\shortauthors{Lima et~al.}

\title [mode = title]{Petal-Graphyne: A Novel 2D Carbon Allotrope for High-Performance Li and Na Ion Storage}

\author[1]{Kleuton A. L. Lima}
\affiliation[1]{
organization={Department of Applied Physics and Center for Computational Engineering and Sciences},
addressline={State University of Campinas}, 
city={Campinas},
postcode={13083-859}, 
state={SP},
country={Brazil}}
\credit{Conceptualization of this study, Methodology, Review and editing, Investigation, Formal analysis, Writing -- review \& editing, Writing -- original draft}

\author[2]{José A. S. Laranjeira}
\affiliation[2]{
organization={Modeling and Molecular Simulation Group},
addressline={São Paulo State University (UNESP), School of Sciences}, 
city={Bauru},
postcode={17033-360}, 
state={SP},
country={Brazil}}
\credit{Conceptualization of this study, Methodology, Review and editing, Investigation, Formal analysis, Writing -- review \& editing, Writing -- original draft}
\author[2]{Nicolas F. Martins}
\credit{Conceptualization of this study, Methodology, Review and editing, Investigation, Formal analysis, Writing -- review \& editing, Writing -- original draft}

\author[3,4]{Alexandre C. Dias}
\affiliation[3]{
organization={Institute of Physics},
addressline={University of Brasília}, 
city={Brasília },
postcode={70910‑900}, 
state={DF},
country={Brazil}}
\affiliation[4]{
organization={International Center of Physics},
addressline={University of Brasília}, 
city={Brasília },
postcode={70910‑900}, 
state={DF},
country={Brazil}}
\credit{Conceptualization of this study, Methodology, Review and editing, Investigation, Formal analysis, Writing -- review \& editing, Writing -- original draft}

\author[2]{Julio R. Sambrano}
\credit{Conceptualization of this study, Methodology, Review and editing, Investigation, Formal analysis, Writing -- review \& editing, Writing -- original draft}

\author[1]{Douglas S. Galvão}
\credit{Conceptualization of this study, Methodology, Review and editing, Investigation, Formal analysis, Writing -- review \& editing, Writing -- original draft}

\author[3,5]{Luiz A. Ribeiro Junior}
\cormark[1]
\cortext[cor1]{Corresponding author}
\affiliation[5]{
organization={Computational Materials Laboratory, LCCMat, Institute of Physics},
addressline={University of Brasília}, 
city={Brasília },
postcode={70910‑900}, 
state={DF},
country={Brazil}}
\credit{Conceptualization of this study, Methodology, Review and editing, Investigation, Formal analysis, Writing -- review \& editing, Writing -- original draft}

\begin{abstract}
 Using density functional theory simulations, this study introduces Petal-Graphyne (PLG), a novel multi-ring metallic structure composed of 4-, 8-, 10-, and 16-membered rings. Its structural, electronic, and lithium/sodium storage properties were comprehensively investigated. PLG exhibits a high theoretical capacity of \SI{1004}{mAh/g} for \ce{Li}, \ce{Na}, and mixed-\ce{Li}/\ce{Na}-ions, surpassing conventional graphite anodes. The material remains metallic, with multiple band crossings at the Fermi level. The optimal energy barriers for \ce{Li} (\SI{0.28}{\electronvolt}) and \ce{Na} (\SI{0.25}{\electronvolt}) on PLG and favorable diffusion coefficients in both monolayer and multilayer configurations are unveiled. The open circuit voltages are \SI{0.47}{\volt} for \ce{Li}, \SI{0.51}{\volt} for \ce{Na}, and \SI{0.54}{\volt} for mixed-ion storage, suggesting stable electrochemical performance. These results highlight PLG as a promising candidate for next-generation lithium and sodium-ion batteries, combining high storage capacity and efficient ion transport.
\end{abstract}




\maketitle

\section{Introduction}

Two-dimensional (2D) carbon allotropes have been extensively studied due to their exceptional structural and electronic properties, which makes them attractive for a variety of applications, such as in energy storage, catalysis, and nanoelectronics \cite{tiwari2016magical,nasir2018carbon,hirsch2010era}. Among these materials, porous 2D carbon allotropes are of particular interest as they present high surface area-to-volume ratio and tunable electronic properties, which are important for efficient ion transport in battery applications \cite{yi2018microporosity,zheng2015two}. Multi-membered rings in such structures can enhance ion diffusion and provide diverse adsorption sites, enhancing their performance in energy storage devices \cite{yi2018microporosity,paul2019carbon,li2021rational}. 

Computational studies, particularly density functional theory (DFT)-based simulations, play a key role in reliably predicting new 2D materials and assessing their stability, electronic properties, and electrochemical performance before experimental realization \cite{LIMA2025116099,you20242d,xiong2024theoretical,martins2024irida}. Examples of already obtained 2D carbon allotropes include graphene \cite{geim2007rise}, graphyne \cite{desyatkin2022scalable}, biphenylene  \cite{fan2021biphenylene}, and monolayer fullerene networks \cite{hou2022synthesis,hudspeth2010electronic,berber2004rigid,baughman1987structure,wallace1947band}. Other examples of predicted/proposed structures are Penta-graphene \cite{zhang2015penta}, Popgraphene \cite{wang2018popgraphene}, Phagraphene \cite{wang2015phagraphene}, Irida-graphene \cite{junior2023irida}, Sun-graphyne \cite{tromer2023mechanical}, and PolyPyGy \cite{LIMA2025116099}, which highlights the diversity of possible carbon allotropes. 

Many 2D carbon allotropes exhibit low conductivity or suboptimal ion diffusion, limiting their performance in lithium and sodium-ion batteries (LIBs and SIBs) \cite{mao2014nanocarbon}. In this way, there is a renewed interest in designing new structures that could overcome these limitations, which require intrinsic metallicity, ensuring efficient charge transport, and offering potential high storage capacities. As an example, high diffusion rates and good storage capacities have been reported for \ce{Li} and \ce{Na} ions on materials such as biphenylene network \cite{chen2023biphenylene, ferguson2017biphenylene}, Irida-graphene \cite{xiong2024theoretical, martins2024irida}, TODD-graphene \cite{santos2024proposing}, Tolanene \cite{ullah2024theoretical}, and TPDH-graphene \cite{gomez2024tpdh}, as well as other related 2D porous carbon materials \cite{liu2024novel,li2023thfs,cheng2021two,li2017psi}.

Inspired by these results, here we proposed a new 2D carbon allotrope named petal-Graphyne (PLG) (Fig.~\ref{fig:system}). PLG has a unique combination of $\text{sp}$ and $\text{sp}^2$ hybridization, featuring 4-, 8-, 10-, and 16-membered rings designed for high-capacity lithium and sodium-ion storage. Based on the recent experimental advancements in the on-surface synthesis of cyclocarbons structures \cite{sun2024surface}, the PLG synthetic feasibility is whithin our present capabilities and could be, in principle, achieved via retro-Bergman ring-opening reactions from molecular precursors containing 4-, 8-, 10-, and 16-membered rings, which are structural motif also present in PLG.

In this work, we have carried out a comprehensive DFT-based study on the structural, electronic, and electrochemical PLG properties. Our results show that PLG has metallic and good lithium and sodium adsorption properties. We have also evaluated its open circuit voltages (OCV). PLG revealed an excellent battery performance for \ce{Li} and \ce{Na}-ions and mixed-ion storage. These results highlight the PLG potential as a next-generation anode material for lithium and sodium-ion batteries, offering a unique combination of high storage capacity, efficient ion transport, and robust electrical conductivity.

\section{Methodology}

PLG structural, electronic, and lithium/sodium storage properties were investigated using DFT simulations with the CASTEP code \cite{clark2005first}. All simulations employed the Perdew-Burke-Ernzerhof (PBE) exchange-correlation functional under the generalized gradient approximation (GGA) scheme \cite{perdew1996generalized}. A plane-wave energy cutoff of \SI{450}{\electronvolt} and a Monkhorst-Pack \textbf{k}-point mesh of $10\times10\times1$ were used for the Brillouin zone sampling. Structural relaxations (atom positions and crystal lattice) were performed with periodic boundary conditions, ensuring minimal residual forces below \SI{E-3}{\electronvolt/\angstrom} and pressure below \SI{0.01}{\giga\pascal} GPa. A vacuum buffer layer of \SI{20}{\angstrom}  was applied along the out-of-plane direction to eliminate spurious mirror cell interactions. Van der Waals corrections were included within the Grimme D2 semi-empirical correction \cite{grimme2006semiempirical} to take into account dispersion effects.

Phonon calculations using density functional perturbation theory (DFPT) \cite{baroni2001phonons} confirmed the PLG dynamical stability. Additionally, \textit{ab initio} molecular dynamics (AIMD) simulations were performed at \SI{1000}{\kelvin} for \SI{5}{\ps}  coupled to a Nosé-Hoover thermostat \cite{nose1984unified} in the NVT ensemble to assess thermal structural resilience, using a $2\times2\times1$ supercell with a $5\times5\times1$ \textbf{k}-points grid. These simulations were monitored for structural integrity, identifying potential bond-breaking regions, atomic reconfigurations, and structural failures.

The electronic band structure and density of states (DOS) were computed using \textbf{k}-points grids of $10\times10\times1$ and $20\times20\times1$, respectively. The metallic nature of X-Graphyne was verified through its band dispersion characteristics. Additionally, adsorption properties were investigated by placing \ce{Li} and \ce{Na} atoms at various randomly chosen adsorption sites and performing full geometry optimizations to determine the most stable configurations using the adsorption locator tool \cite{metropolis1953equation,kirkpatrick1983optimization,vcerny1985thermodynamical}. 

The adsorption energy ($E_\text{ads}$) was calculated as: $E_{\text{ads}} = E_{\text{X-Graphyne+M}} - \left( E_{\text{X-Graphyne}} + E_{\text{M}} \right)$, where $E_{\text{X-Graphyne+M}}$ is the PLG total energy with an adsorbed ion (\ce{M} = \ce{Li} or \ce{Na}), $E_{\text{X-Graphyne}}$ is the energy of pristine PLG, and $E_{\text{M}}$ is the energy of an isolated metal ion. The nudged elastic band (NEB) method \cite{makri2019preconditioning, barzilai1988two, bitzek2006structural} was used to evaluate \ce{Li} and \ce{Na} diffusion barriers, providing insights into ion mobility across the surface.

\section{Results}

\subsection{Structural and Mechanical Properties}

\begin{figure}[pos=!htb]
    \centering
    \includegraphics[width=\linewidth]{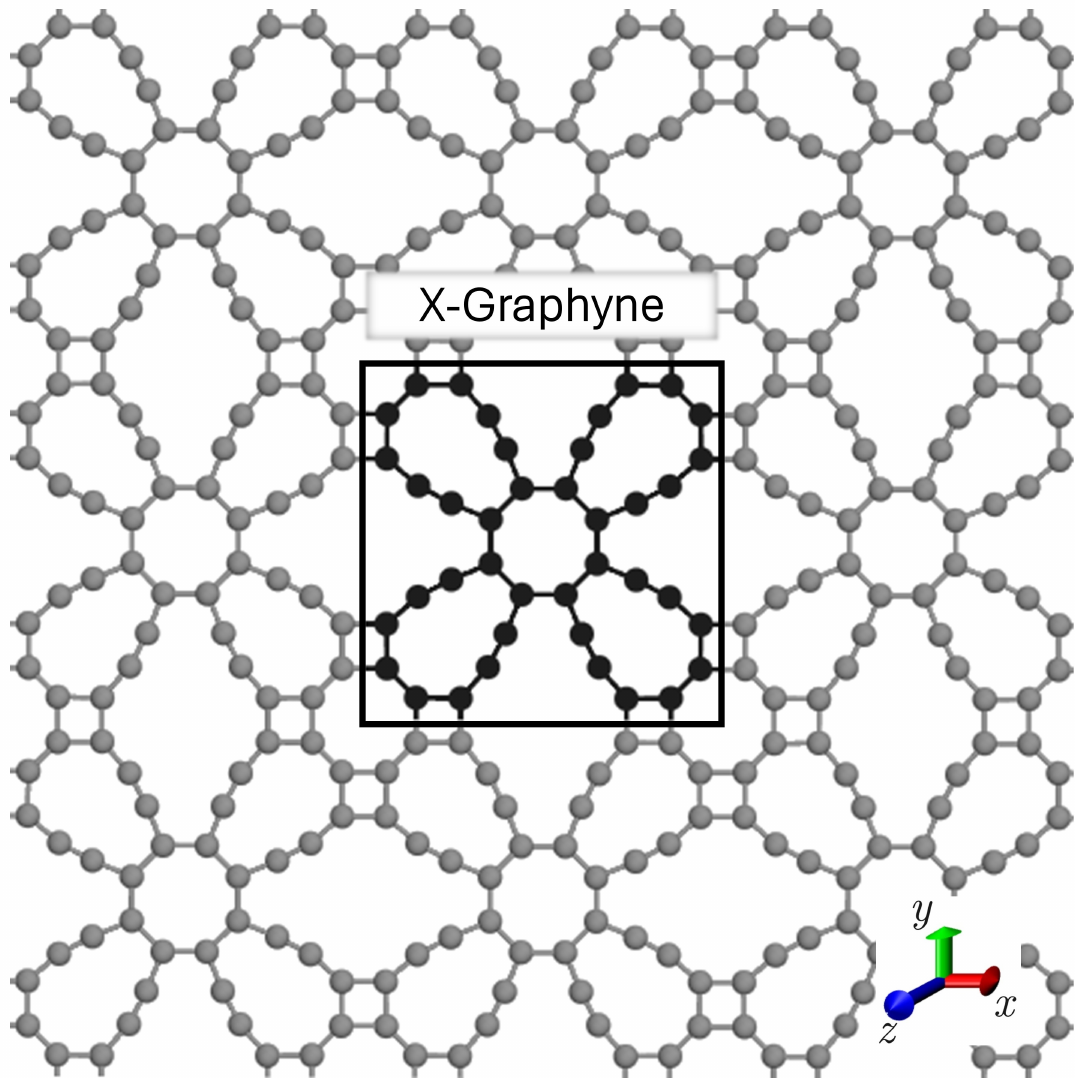}
    \caption{Atomic structure of petal-like graphyne (PLG). The black box highlights the structural unit cell, which contains $40$ carbon atoms arranged in a porous framework with 4-, 8-, 10-, and 12-membered rings.}
    \label{fig:system}
\end{figure}

First, we present the PLG structural characteristics. Fig.~\ref{fig:system} illustrates the atomic structure of this novel material, which is composed of 4-, 8-, 10, and 16-membered rings arranged in a periodic lattice. The unit cell (black box) comprises $40$ carbon atoms and exhibits a porous topology. The symmetry group PM (CS-1) describes the material with an IT number of $6$, reflecting its low symmetry and anisotropic atomic arrangement. The optimized unit cell parameters are $a=$\SI{11.69}{\angstrom} and $b=$\SI{11.68}{\angstrom}. PLG has formation energy of \SI{-7.92}{\electronvolt}, in the range value of other 2D carbon allotropes,  which suggests thermodynamic stability.

\begin{figure}[pos=!htb]
    \centering
    \includegraphics[width=\linewidth]{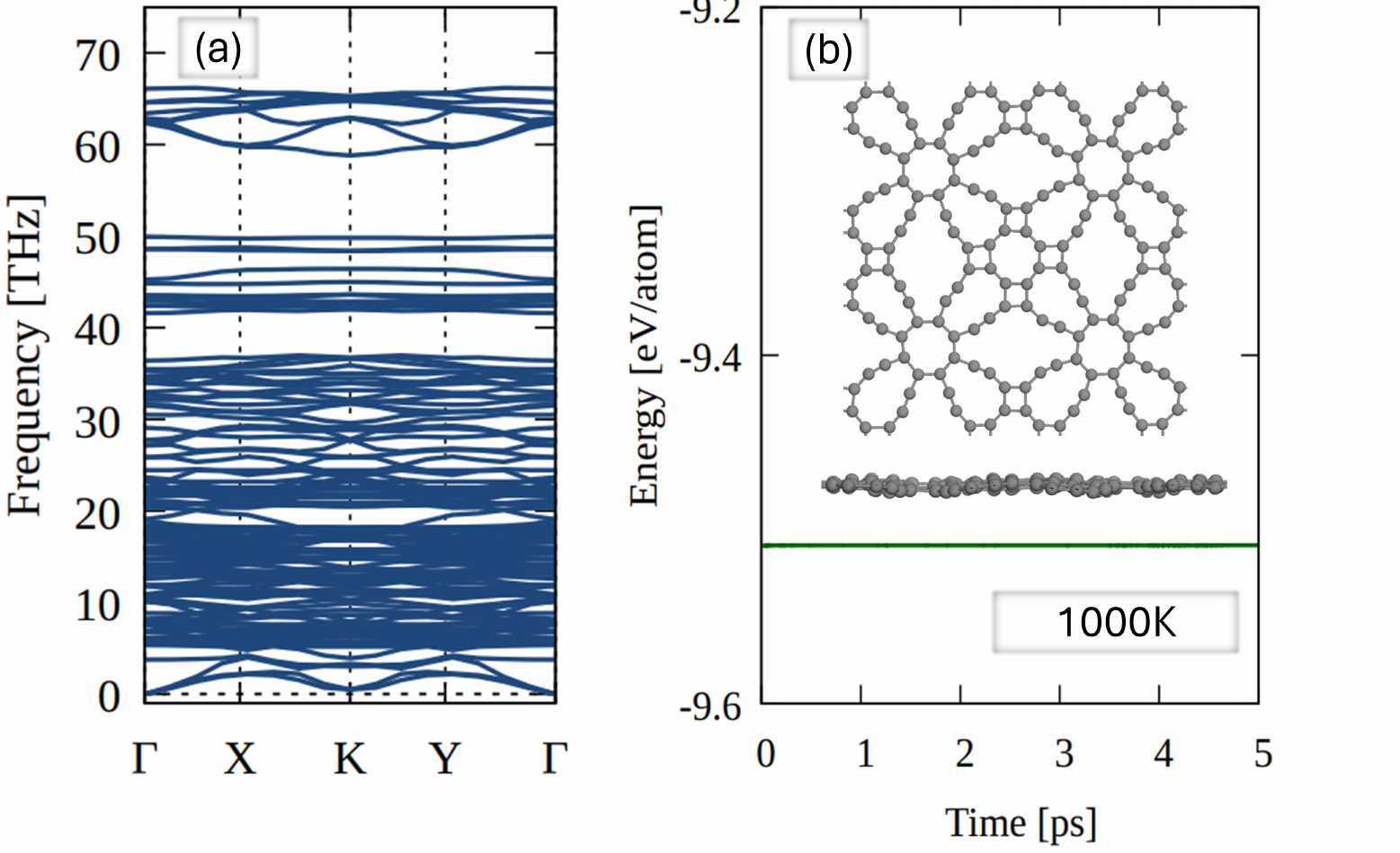}
    \caption{PLG stability analyses. (a) Phonon dispersion relation along high-symmetry points in the Brillouin zone and (b) AIMD simulation results at \SI{1000}{\kelvin} over \SI{5}{\ps}. The inset panels display the structure's final top and side views at \SI{5}{ps}.}
    \label{fig:phonons-aimd}
\end{figure}

Phonon dispersion and AIMD simulations were performed to evaluate the PLG structural stability and AIMD), as shown in Fig.~\ref{fig:phonons-aimd}. The phonon dispersion is shown in Fig.~\ref{fig:phonons-aimd}(a) and confirms the PLG dynamical stability, as no imaginary frequencies are observed across the Brillouin zone. Three acoustic modes corresponding to longitudinal, transverse, and out-of-plane vibrations were also observed. Combining $\text{sp}$ and $\text{sp}^2$ hybridizations in the 2D lattice results in inhomogeneous bonding strengths. Regions with $\text{sp}$-hybridized bonds create local strain, stiffening specific vibrational modes, while $\text{sp}^2$-hybridized regions allow for softer modes at lower frequencies. This variation in bond strength leads to gaps in the phonon spectrum.

AIMD simulations carried out at \SI{1000}{\kelvin} for \SI{5}{\ps} (Fig.~\ref{fig:phonons-aimd}(b)) confirm PLG thermal stability. The AIMD energy profile remains almost constant, indicating the absence of significant structural distortions or atomic reconstructions. The inset panels display the top and side views of the final AIMD snapshot, revealing that a planar-like structure is preserved even at high temperatures. No bond breaking or substantial out-of-plane deformations were observed, further confirming the PLG thermal robustness.

\begin{figure}[pos=!htb]
    \centering
    \includegraphics[width=\linewidth]{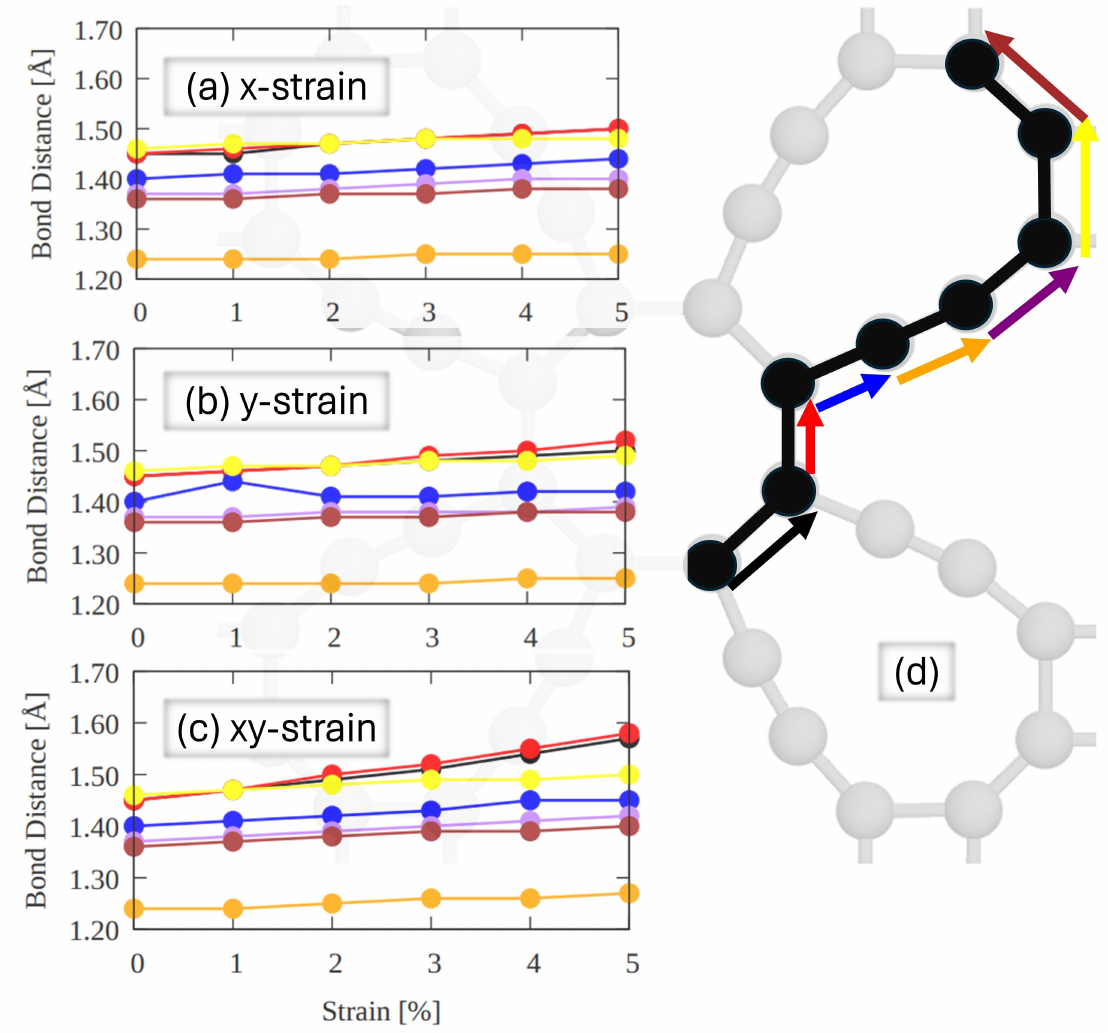}
    \caption{PLG bond length values evolution under applied strain. (a) $x$-strain, (b) $y$-strain, and (c) $xy$-strain show the variation in bond distances as a function of strain percentage, highlighting the anisotropic mechanical response. (d) Structural representation of the key bonds, with arrows indicating the primary deformation pathways under different strain conditions. }
    \label{fig:bondlengths}
\end{figure}

To better understand the PLG mechanical resilience, the evolution of bond lengths under different strain conditions was analyzed, as shown in Fig.~\ref{fig:bondlengths}. The subfigures \ref{fig:bondlengths}(a), \ref{fig:bondlengths}(b), and \ref{fig:bondlengths}(c) illustrate the bond length variations as a function of the applied $x$-strain, $y$-strain, and $xy$-strain, respectively. Fig.~\ref{fig:bondlengths}(d) provides a structural representation of the key bonds, with different colors indicating their behavior under deformation.

The bond length changes under strain exhibit an anisotropic behavior, which is expected given the PLG multi-ringed topology. Under $x$-strain (Fig.~\ref{fig:bondlengths}(a)), the bonds oriented along the stretching direction experience a gradual increase in length, while those perpendicular to it remain relatively unchanged. This feature indicates that PLG can accommodate stress by elongating specific bonds rather than uniformly distributing the strain. Similarly, under the $y$-strain (Fig.~\ref{fig:bondlengths}(b)), the bond length variations follow a similar trend but with a distinct set of bonds experiencing elongation. The observed differences between $x$-strain and $y$-strain responses suggest directional-dependent mechanical properties, which could be important in practical applications requiring strain engineering.

When subjected to $xy$-strain (Fig.~\ref{fig:bondlengths}(c)), the bond deformation patterns become more complex due to shear stress interactions, leading to a more uniform increase in bond lengths across the different orientations. The structural representation in Fig.~\ref{fig:bondlengths}(d) further highlights how specific bonds stretch in different directions, with colored arrows indicating the primary deformation pathways. Both short and long bonds in the PLG framework result in a nontrivial stress redistribution, which may enhance its fracture resistance and mechanical flexibility.

The PLG mechanical properties were evaluated by calculating the directional Young's modulus \( Y(\theta) \) and Poisson's ratio \( \nu(\theta) \) values, as presented in Fig.~\ref{fig:mechprop}. These quantities were obtained using the elastic constants and 
the equations presented below. The estimated elastic constants 
(\( C_{11} = 144.78 \) \si{\newton/m}, \( C_{12} = 111.80 \) \si{\newton/m}, \( C_{22} = 144.68 \) \si{\newton/m}, and \( C_{66} = 61.64 \) \si{\newton/m}) satisfy the Born–Huang mechanical stability criteria of the orthorhombic crystal ($C_{11}C_{22}-C^{2}_{12}>0$ and $C_{44}>0$ ) \cite{PhysRevB.90.224104,doi:10.1021/acs.jpcc.9b09593}, confirming the PLG structural robustness. The Young’s modulus \( Y(\theta) \) is given by:

\begin{equation}
Y(\theta) = \frac{C_{11}C_{22} - C_{12}^{2}}{C_{11} \sin^{4} \theta + 
\left( \frac{C_{11}C_{22} - C_{12}^{2} - 2C_{12}}{C_{66}} \right) 
\cos^{2} \theta \sin^{2} \theta + C_{22} \cos^{4} \theta },
\end{equation}

\noindent and the Poisson’s ratio \( \nu(\theta) \) is given by:

\begin{equation}
\nu(\theta) = \frac{C_{12} (\cos^{4} \theta + \sin^{4} \theta) - 
\left( \frac{C_{11} + C_{22} - \frac{C_{11}C_{22} - C_{12}^{2}}{C_{66}}}{C_{66}} \right) 
\cos^{2} \theta \sin^{2} \theta }{C_{11} \sin^{4} \theta + 
\left( \frac{C_{11}C_{22} - C_{12}^{2} - 2C_{12}}{C_{66}} \right) 
\cos^{2} \theta \sin^{2} \theta + C_{22} \cos^{4} \theta }.
\end{equation}

In Fig.~\ref{fig:mechprop}, we can observe that the PLG calculated Young's modulus reaches a maximum value of approximately \SI{180}{\newton/m} and a minimum one of around \SI{60}{\newton/m}, highlighting its significant mechanical anisotropy. Compared to graphynes, which exhibit Young's modulus values ranging from \SIrange{180}{220}{\newton/m} \cite{shao2012temperature}, PLG exhibits a lower overall stiffness due to its larger multi-membered rings that introduce additional structural flexibility. Similarly, compared with graphdiyne, which has a modulus of \SI{100}{\newton/m}–\SI{150}{\newton/m}  \cite{pei2012mechanical}, PLG presents a comparable but slightly wider range of mechanical responses, indicating that its porous structure enhances flexibility while maintaining sufficient rigidity for structural stability. 

\begin{figure*}[pos=!htb]
    \centering
    \includegraphics[width=\linewidth]{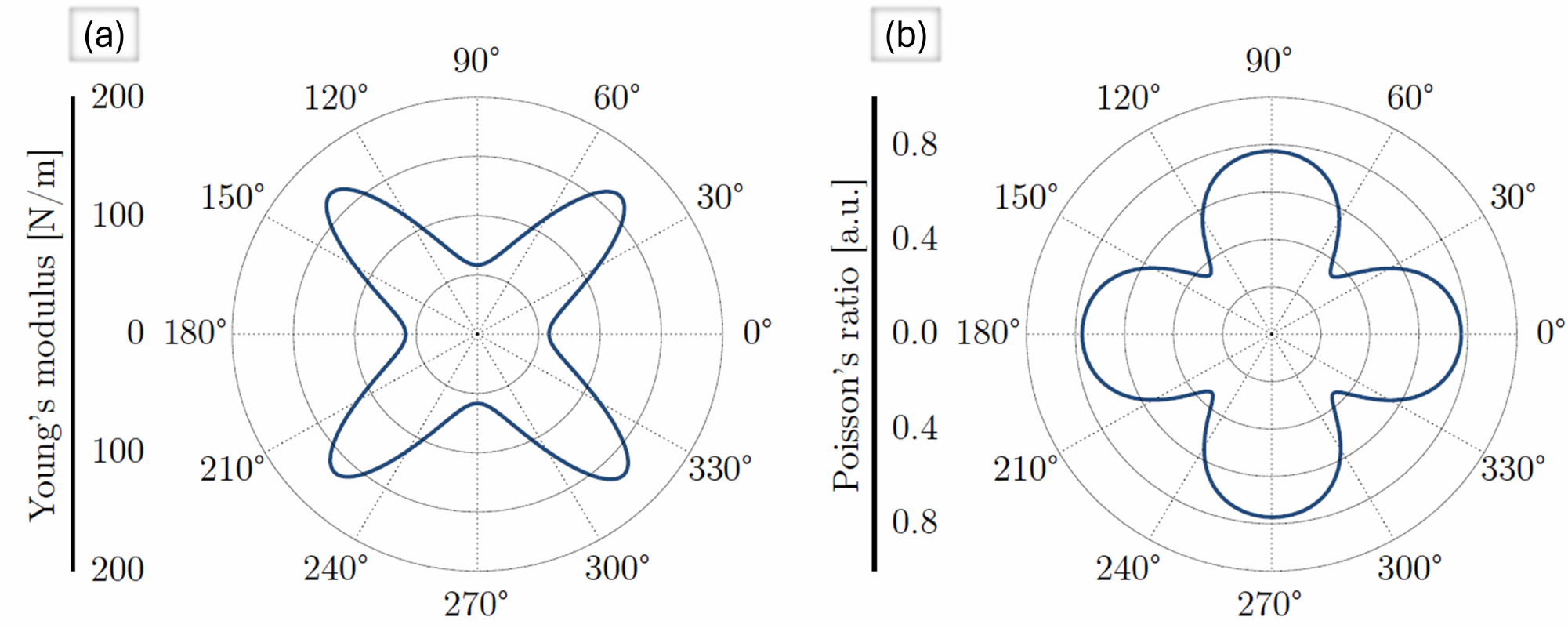}
    \caption{PLG anisotropic mechanical properties. (a) The polar plots of Young's modulus and (b) Poisson's ratio values.}
    \label{fig:mechprop}
\end{figure*}

The Poisson's ratio distribution is shown in Fig;~\ref{fig:mechprop}(b), quantifying the material's lateral response to applied strain. The plot exhibits a fourfold symmetry, exhibiting the expected direction-dependent deformation behavior. The maximum Poisson's ratio reaches approximately $0.8$, which is relatively high for 2D materials, while remaining strictly positive in all directions, confirming that PLG is not auxetic. Instead, the material follows the conventional Poisson effect, where stretching it along one direction leads to contraction in the perpendicular direction. The variation in Poisson's ratio across different orientations suggests that PLG undergoes non-uniform lateral deformation as the result of the multi-ringed topology and anisotropic bonding environment. 

\subsection{Electronic and Optical Properties}

\begin{figure}[pos=!htb]
    \centering
    \includegraphics[width=\linewidth]{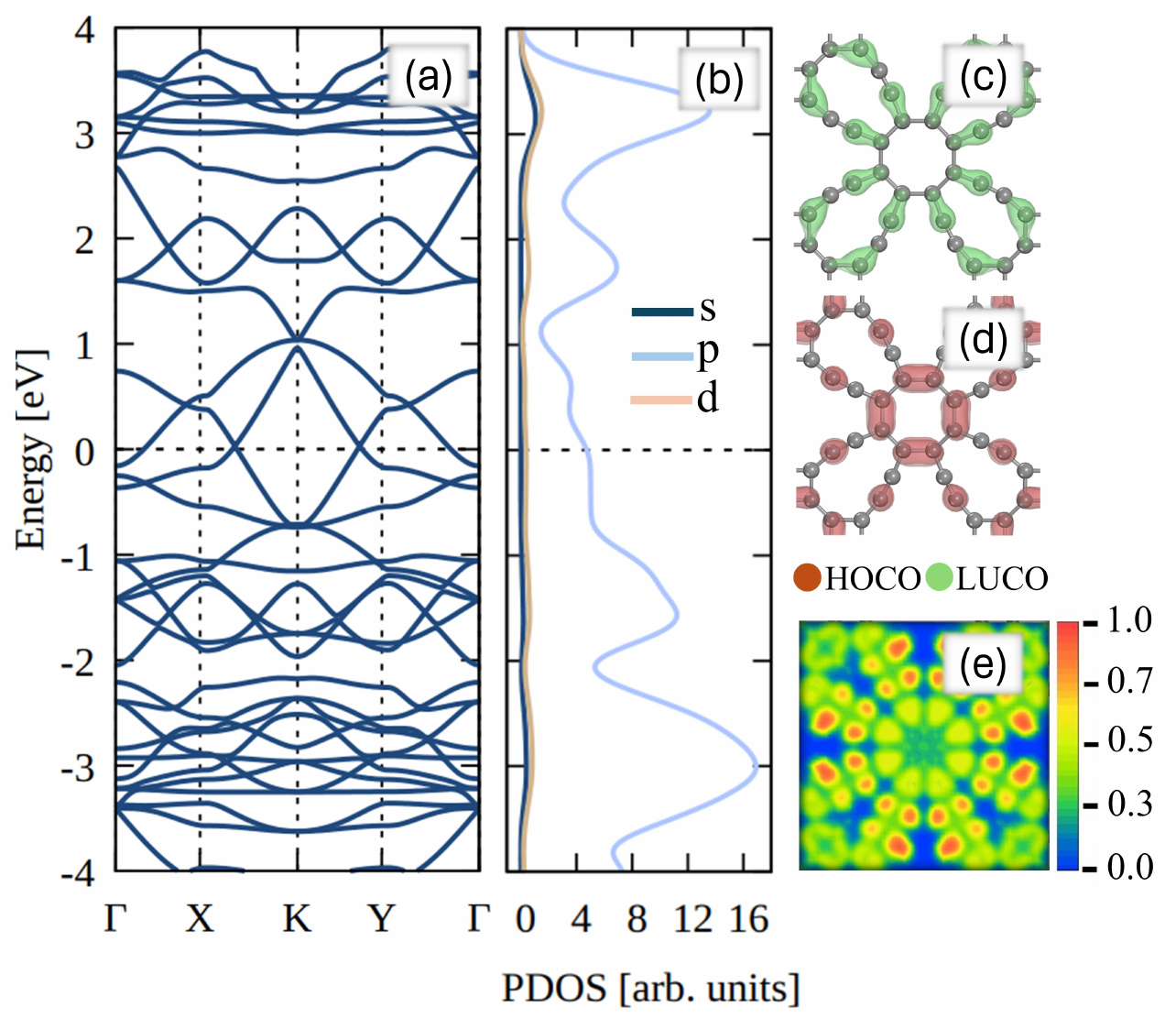}
    \caption{PLG electronic properties. (a) Electronic band structure, (b) Corresponding projected density of states (PDOS), (c) Highest occupied crystalline orbital (HOCO), (d) Lowest unoccupied crystalline orbital (LUCO), and (e) Electron localization function.}
    \label{fig:elecprop}
\end{figure}

In Fig.~\ref{fig:elecprop}, we present the PLG electronic band structure \ref{fig:elecprop}(a), the corresponding projected density of states (PDOS) \ref{fig:elecprop}(b), the highest occupied crystalline orbital (HOCO) \ref{fig:elecprop}(c), the lowest unoccupied crystalline orbital (LUCO) \ref{fig:elecprop}(d), and the electron localization function \ref{fig:elecprop}(e).

Fig.~\ref{fig:elecprop}(a) displays the PLG electronic band structure, confirming that it is metallic, as multiple bands cross the Fermi level (set at \SI{0}{\electronvolt}). Unlike gamma-graphyne, which exhibits a semiconducting behavior with an electronic band gap in the range of \SIrange{0.4}{1.0}{\electronvolt}\cite{aliev2025planar}, and graphdiyne, which has a larger band gap of approximately \SI{0.5}{}–\SI{1.2}{\electronvolt} \cite{koo2014widely}, PLG maintains a gapless electronic structure. The PLG metallicity is a direct consequence of its extended $\pi$-conjugation and multi-ring topology, where the delocalized electron network provides high electrical conductivity, a key feature for applications in energy storage and electronic devices.

The PDOS in Fig.~\ref{fig:elecprop}(b) shows that the electronic states near the Fermi level are predominantly composed of $2pz$-orbitals. This feature indicates that the $\pi$-electron network dominates its electronic properties. Importantly, this behavior is similar to that observed in graphyne and graphdiyne, where the sp$^2$-hybridized carbon atoms contribute significantly to the density of states. However, in PLG, the higher density of states at the Fermi level suggests an enhanced charge carrier concentration, making it a superior candidate for electronic transport applications compared to its semiconducting counterparts.

Figs.~\ref{fig:elecprop}(c) and \ref{fig:elecprop}(d) show the HOCO and LUCO orbital distributions, respectively. The HOCO and LUCO  contributions are widely distributed throughout the framework. This delocalization ensures efficient charge transport, which is beneficial for electrode materials in batteries and supercapacitors. In contrast, graphyne and graphdiyne often exhibit localized electronic states due to their larger ring sizes, which can limit carrier mobility \cite{long2011electronic,li2020structural,li2023artificial}.

The electron localization function (Fig.~\ref{fig:elecprop}(e)) shows a uniform charge distribution throughout the structure, further reinforcing its metallic nature. The high electron density in the 10-membered ring indicates that charge carriers can move freely, enhancing the overall electrical conductivity. Again, this behavior differs from graphyne and graphdiyne \cite{sha2021first,muz2020electronic}, where the electron density is often concentrated around specific bonds due to localized conjugation effects.

\begin{figure}[pos=!htb]
    \centering
    \includegraphics[width=\linewidth]{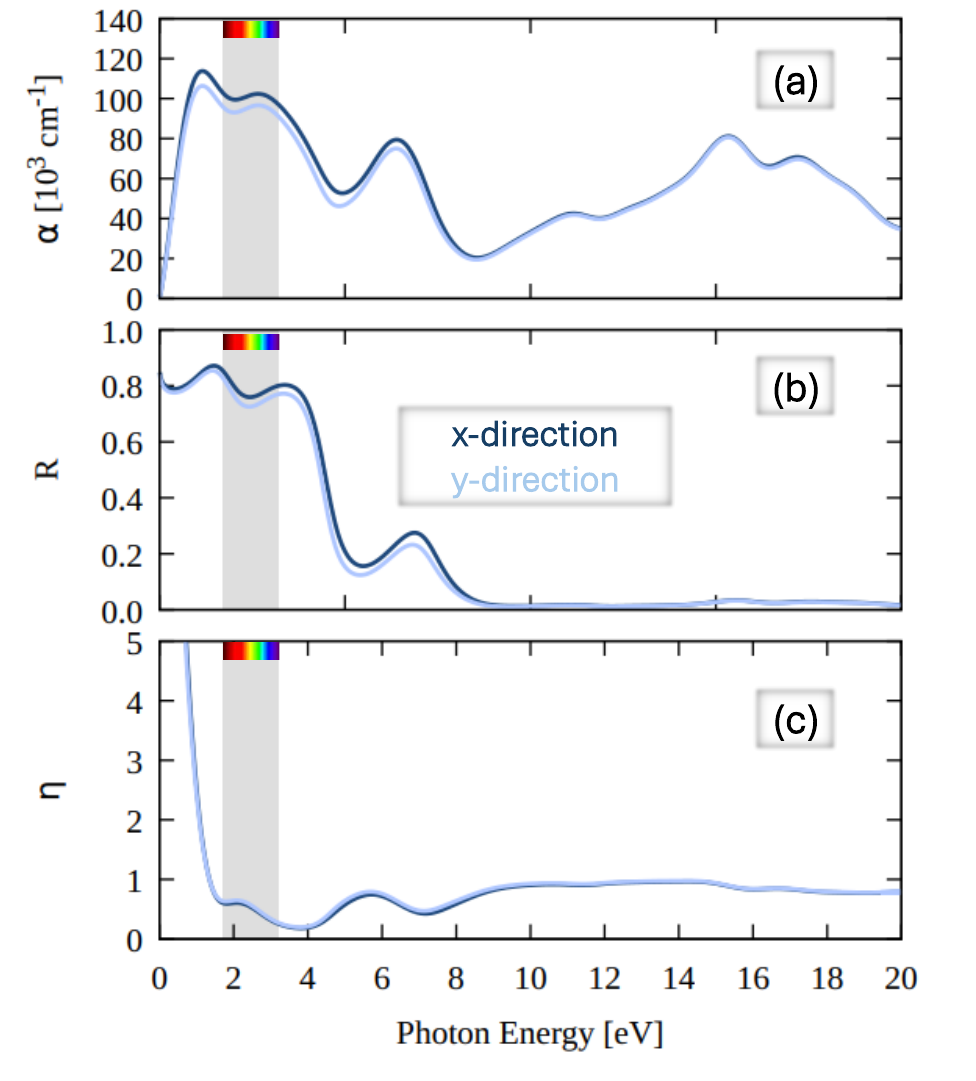}
    \caption{PLG optical properties as a function of photon energy. (a) Absorption coefficient ($\alpha$), (b) reflectivity ($R$) spectrum, and (c) refractive index ($\eta$).}
    \label{fig:opticalprop}
\end{figure}

To investigate the PLG potential for optoelectronic applications, the optical absorption coefficient ($\alpha$), reflectivity ($R$), and refractive index ($\eta$) as functions of photon energy were examined, as shown in Fig.~\ref{fig:opticalprop}. 

Fig.~\ref{fig:opticalprop}(a) shows the absorption coefficient $\alpha$. The absorption spectrum exhibits a strong and broad response, with activities in the infrared (IR), visible (VIS), and ultraviolet (UV) regions, with peaks also extending into these regions. PLG exhibits high absorption above \SI{2}{\electronvolt}, indicating that it can interact efficiently with UV and visible light. Compared to graphyne and graphdiyne, which exhibit absorption edges around \SI{1.5}{\electronvolt}–\SI{2.0}{\electronvolt} \cite{hou2018study,ge2018review}, PLG presents a blue-shifted absorption profile, suggesting an enhanced optical response in higher-energy regions. The anisotropy between the $x$- and $y$-directions is slightly different, indicating nearly isotropic optical absorption.

The reflectivity spectrum (R), which describes the fraction of incident light that is reflected by the material, is presented in Fig.~\ref{fig:opticalprop}(b). PLG exhibits a high reflectivity across the solar emission spectral range (\SIrange{0}{4}{\electronvolt}), particularly in the visible region, making it an efficient light-reflecting material with reflectivity around \SI{80}{\percent}. The reflectivity sharply drops after \SI{3}{\electronvolt}. Compared to graphyne and graphdiyne, which exhibit moderate reflectivity variations \cite{hou2018study,ge2018review,jafarzadeh2020electronic}, PLG shows a higher uniform and consistently high reflectivity in the infrared and visible regions of the electromagnetic spectrum.

Fig.~\ref{fig:opticalprop}(c) illustrates the refractive index ($\eta$), which determines the phase velocity of light within the material. PLG exhibits a peak in the refractive index at low photon excitation energies, followed by a gradual decline, characteristic of materials with metallic behavior. This trend contrasts with semiconducting graphyne and graphdiyne, which display higher refractive indices in the visible range due to their band gap-related optical transitions \cite{hou2018study,ge2018review,jafarzadeh2020electronic}. 

\subsection{Li/Na-ion diffusion on monolayer and bilayer PLG}

\begin{figure*}[pos=!htb]
    \centering
    \includegraphics[width=\linewidth]{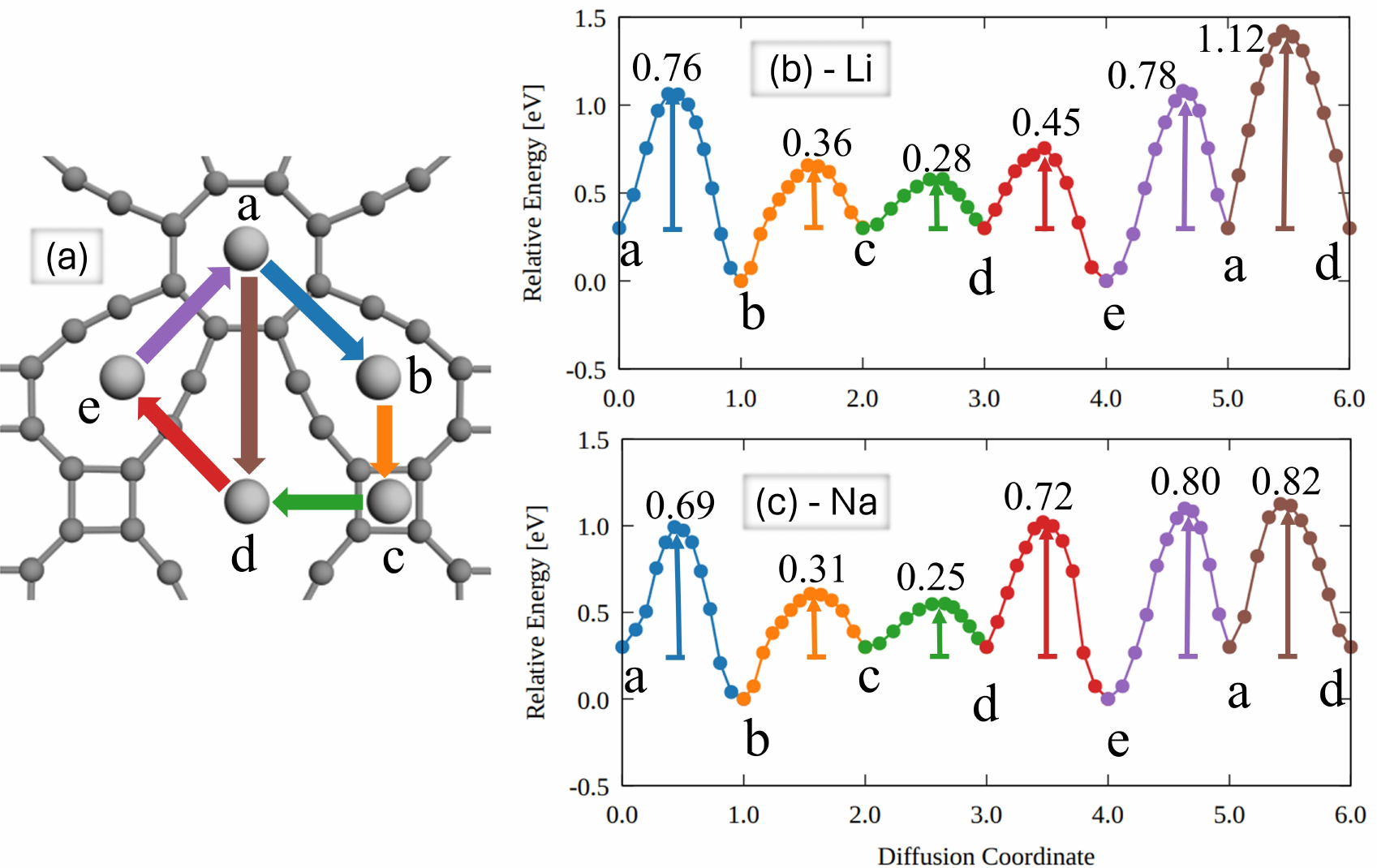}
    \caption{PLG ion diffusion pathways and energy barriers. (a) Schematic representation of \ce{Li}/\ce{Na} diffusion paths between adsorption sites (a–e) within the porous PLG structure. (b) The diffusion energy profile for \ce{Li}-ions shows the lowest energy barrier of \SI{0.28}{\electronvolt}  along the c–d path and the highest barrier of \SI{1.12}{\electronvolt} along the d–a transition. (c) The diffusion energy profile for \ce{Na}-ions has a minimum barrier of \SI{0.25}{\electronvolt} and a maximum of \SI{0.82}{\electronvolt}.}
    \label{fig:barriers}
\end{figure*}

Fig.~\ref{fig:barriers} shows the diffusion pathways and energy barriers for \ce{Li} and \ce{Na} ions analyzed and illustrated to evaluate PLG ion storage and transport capabilities. Diffusion behavior plays a crucial role in determining the efficiency of a material as an anode for lithium-ion batteries (LIBs) and sodium-ion batteries (SIBs). Low diffusion barriers lead to faster charge/discharge rates and improved electrochemical performance.

Fig.~\ref{fig:barriers}(a) presents possible diffusion pathways for \ce{Li} and \ce{Na} ions within the PLG lattice, highlighting key adsorption sites (a, b, c, d, and e) and their respective transition routes. The pathways span the multi-membered rings, allowing ion migration through interconnected pores in the structure. The diffusion energy profiles for \ce{Li} and \ce{Na} ions are shown in Fig.~\ref{fig:barriers}(b) and \ref{fig:barriers}(c), respectively, where the relative energy barriers along different pathways are presented. For \ce{Li}-ion diffusion (Fig.~\ref{fig:barriers}(b)), the lowest energy barrier is \SI{0.28}{\electronvolt}, observed along the c–d path, while the highest barrier reaches \SI{1.12}{\electronvolt} along the d–a transition. The moderate energy barriers in multiple pathways suggest that \ce{Li} ions can efficiently migrate through the structure, though some transitions require additional energy input (energy-activated process).

For \ce{Na}-ion diffusion (Fig.~\ref{fig:barriers}(c)), the lowest energy barrier is \SI{0.25}{\electronvolt}, slightly lower than \ce{Li} ions, indicating that \ce{Na} ions can migrate with similar or improved mobility along specific paths. However, the highest \ce{Na} diffusion barrier reaches \SI{0.82}{\electronvolt}, which, although lower than the highest \ce{Li} barrier, suggests that some migration pathways may still be limited. Even with distinct ion barriers, both \ce{Li} and \ce{Na} ions prefer to migrate across the 4-membered and 16-membered rings. The results suggest that PLG has comparable \ce{Li}/\ce{Na} ion migrations like other similar structures, such as graphyne (x/y eV), graphdiyne (x/y eV), irida-graphene (0.19/0.09 \si{\electronvolt}) \cite{xiong2024theoretical, martins2024irida}, Biphenylene (0.23/0.20 \si{\electronvolt})\cite{duhan20232, han2022biphenylene} and T-graphene (0.37/0.35 \si{\electronvolt}) \cite{hu2020theoretical}.

\begin{figure}[pos=!htb]
    \centering
    \includegraphics[width=\linewidth]{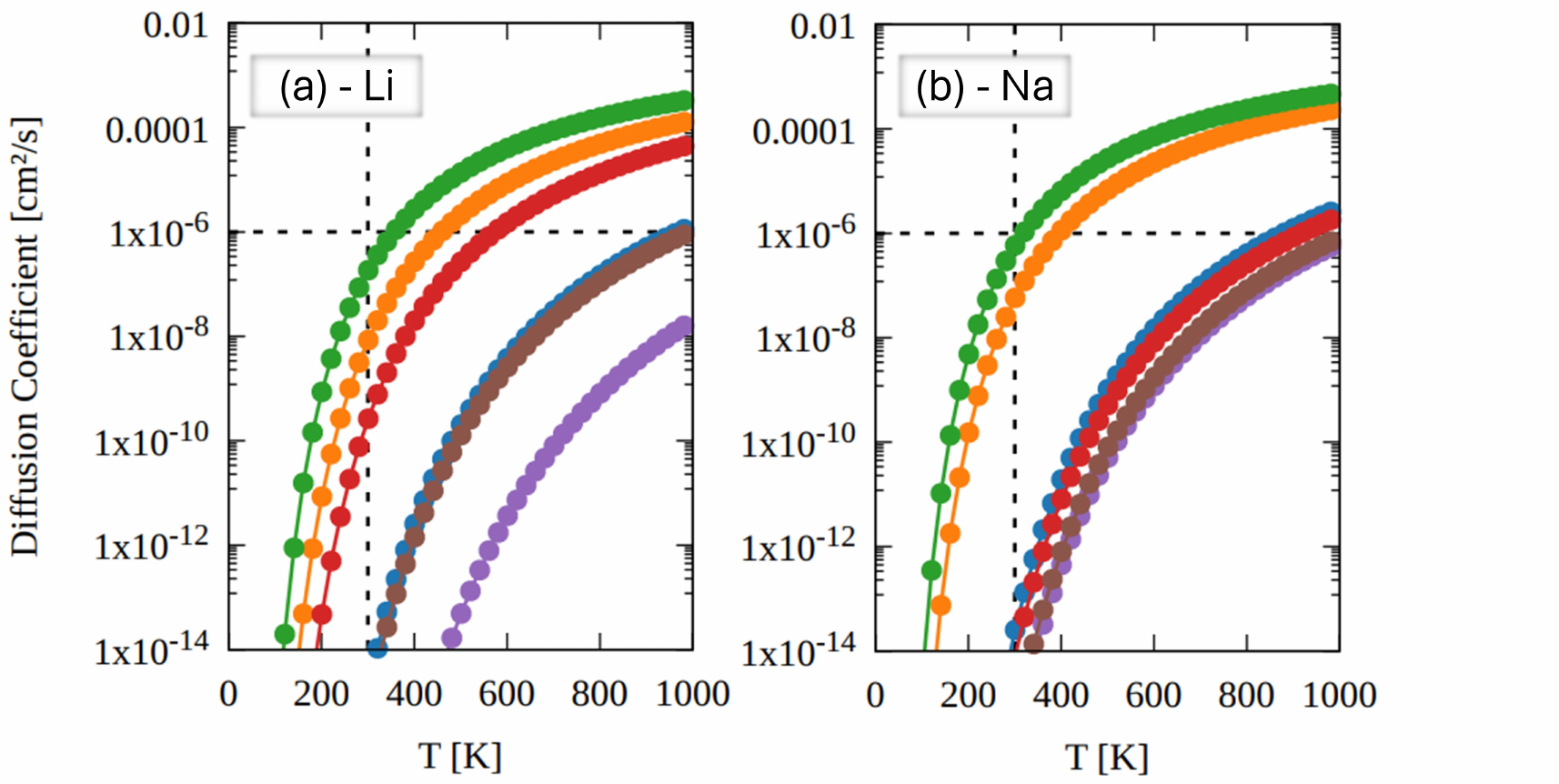}
    \caption{Temperature dependence of the diffusion coefficients for \ce{Li} and \ce{Na} ions. (a) Diffusion coefficient of \ce{Li} ions and (b) diffusion coefficient of \ce{Na} ions. The black dashed line refers to the graphene diffusion coefficient at room temperature.}
    \label{fig:diffusion}
\end{figure}

To further analyze the PLG ionic transport properties, the diffusion coefficients for \ce{Li} and \ce{Na} ions as a function of temperature were calculated and presented in Fig.~\ref{fig:diffusion}. The diffusion coefficient \(D_{\text{coeff}}(T) \) was determined using the Arrhenius-type equation:

\begin{equation}
D_{\text{coeff}}(T) = L^2 \nu_0 \exp \left( -\frac{\Delta E_b}{k_B T} \right),
\end{equation}

\noindent where \( \Delta E_b \) represents the diffusion barrier, \( L \) is the length of the diffusion path, \( T \) is the absolute temperature, \( \nu_0 \) is the vibrational frequency (typically \SI{10}{\tera\hertz}), and \( k_B \) is the Boltzmann constant (\( 8.62 \times 10^{-5} \) \si{\electronvolt/\kelvin}) \cite{gao2023twin}.

Fig.~\ref{fig:diffusion}(a) presents the diffusion coefficients for \ce{Li} ions, while Fig.~\ref{fig:diffusion}(b) shows the values for \ce{Na} ions. At room temperature, \ce{Li} exhibits a diffusion coefficient of approximately \SI{E-8}{\cm^{2}/s}, while \ce{Na} shows a slightly lower value. As expected, both ions exhibit exponential increases in diffusion coefficients with increasing temperature, consistent with the Arrhenius relation. The black dashed line refers to the graphene diffusion coefficient at ambient conditions for comparison.

Comparing the results with graphyne and graphdiyne, PLG exhibits similar or slightly enhanced ionic mobility, particularly at higher temperatures \cite{zhang2013graphdiyne,bartolomei2024sodium,lemaalem2023graphyne}. The low diffusion barriers observed in specific pathways (Fig.~\ref{fig:barriers}) contribute to the improved transport properties. Thus, the PLG high diffusivity can efficiently support fast charge/discharge cycles, reinforcing its potential as a promising anode material for LIBs and SIBs.

\begin{figure}
    \centering
    \includegraphics[width=\linewidth]{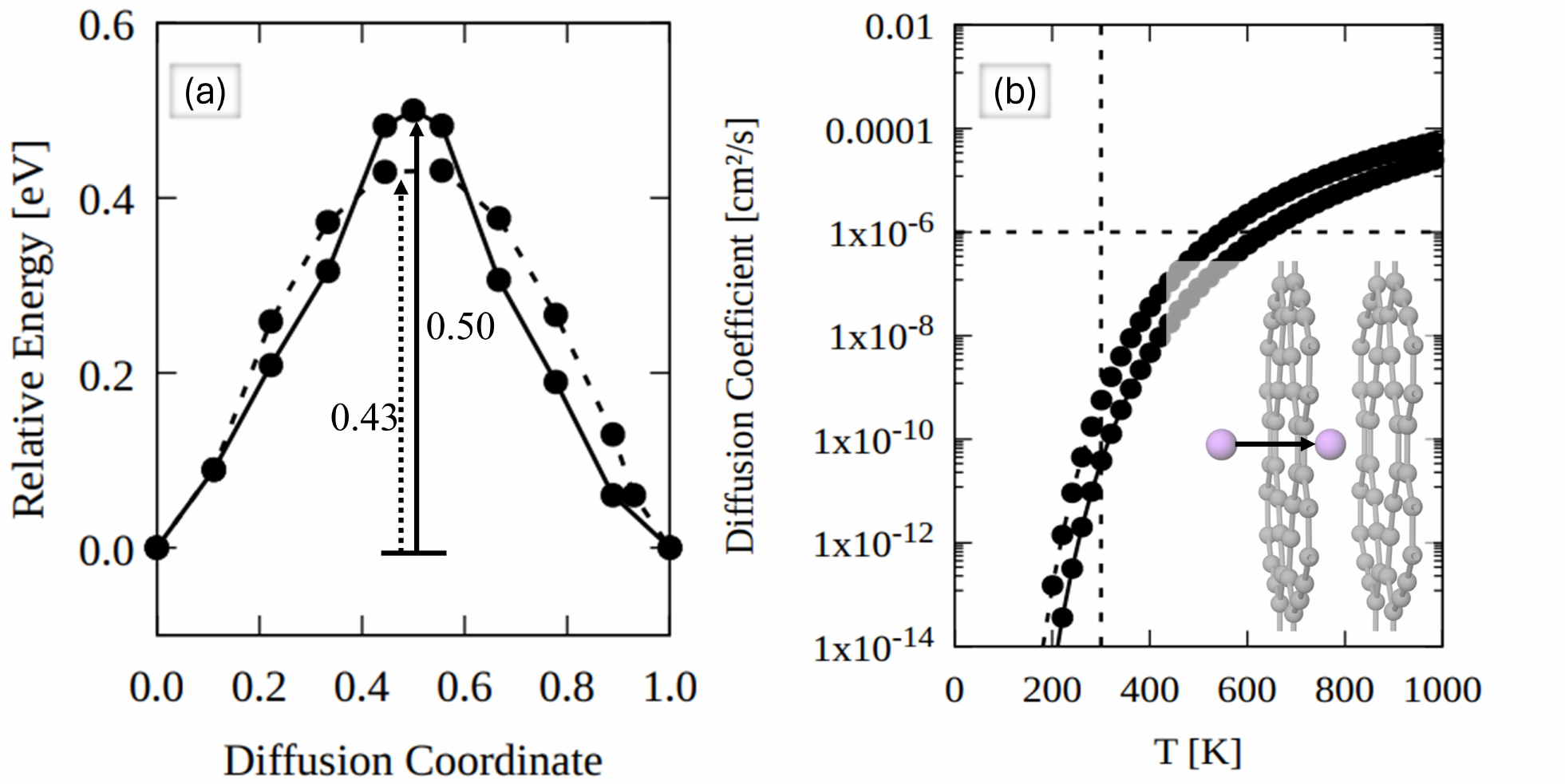}
    \caption{PLG interlayer diffusion of \ce{Li} and \ce{Na} ions. (a) Energy profile for ion migration between two adjacent PLG layers. (b) Temperature-dependent diffusion coefficient for interlayer migration. The inset illustrates the interlayer hopping mechanism, where ions move through the pores between two PLG layers.}
    \label{fig:2diffusion}
\end{figure}

To evaluate the feasibility of \ce{Li} and \ce{Na} ion migration across multiple PLG layers, the vertical diffusion pathway between two adjacent layers (bilayer) was analyzed, as illustrated in Fig.~\ref{fig:2diffusion}. Understanding interlayer diffusion is crucial for assessing the PLG viability as an anode material in stacked electrode configurations, which can enhance the energy storage capacity of battery systems. 

Fig.~\ref{fig:2diffusion}(a) presents the diffusion energy profile for ion migration between two PLG layers. The calculated diffusion barrier for \ce{Li} ions is approximately \SI{0.50}{\electronvolt}, while for \ce{Na} ions, it is \SI{0.43}{\electronvolt}. These values indicate that interlayer diffusion is energetically feasible, albeit with slightly higher barriers than in-plane migration (Fig.~\ref{fig:barriers}). The higher interlayer barrier can be attributed to the van der Waals interactions between the PLG layers, which introduce additional resistance to ion movement. 

The temperature-dependent diffusion coefficient for interlayer migration, calculated using the Arrhenius relation, is presented in Fig.~\ref{fig:2diffusion}(b). The inset illustrates the interlayer hopping mechanism, where ions move through the pores between two PLG layers. At room temperature, the diffusion coefficient for \ce{Li} ions remains in the range of \SI{E-10}{\cm^{2}/s}, which is lower than the in-plane diffusion but still suitable for battery applications \cite{lee2022understanding}. The inset schematic illustrates the interlayer hopping mechanism, where \ce{Li} or \ce{Na} ions move through the pores between two PLG layers. Compared to graphyne and graphdiyne, which typically exhibit higher interlayer barriers due to their denser frameworks \cite{li2014graphdiyne}, PLG presents a more accessible interlayer diffusion path, suggesting its potential for multilayer electrode architectures.

\subsection{Storage Capacity and Open-circuit Voltage}

Metal adsorption analyses were carried out to further assess the practical application of PLG as an anode material for LIBs and SIBs. Different adsorption sites on the PLG surface were systematically examined for single \ce{Li}/\ce{Na} adsorption. Our results indicate that the pore regions on the PLG lattice exhibit the highest stability for starting the lithiation/sodiation processes. \ce{Li} and \ce{Na} metals were selectively introduced to avoid clustering during charge/discharge cycles while ensuring appropriate \ce{Li}-\ce{Li} and \ce{Na}-\ce{Na} distances, as observed in the bulk phase. Consequently, the maximum metal loading is achieved when $18$ \ce{Li} and $18$ \ce{Na} atoms are adsorbed onto the PLG monolayer.

\begin{figure*}[pos=!htb]
    \centering
    \includegraphics[width=\linewidth]{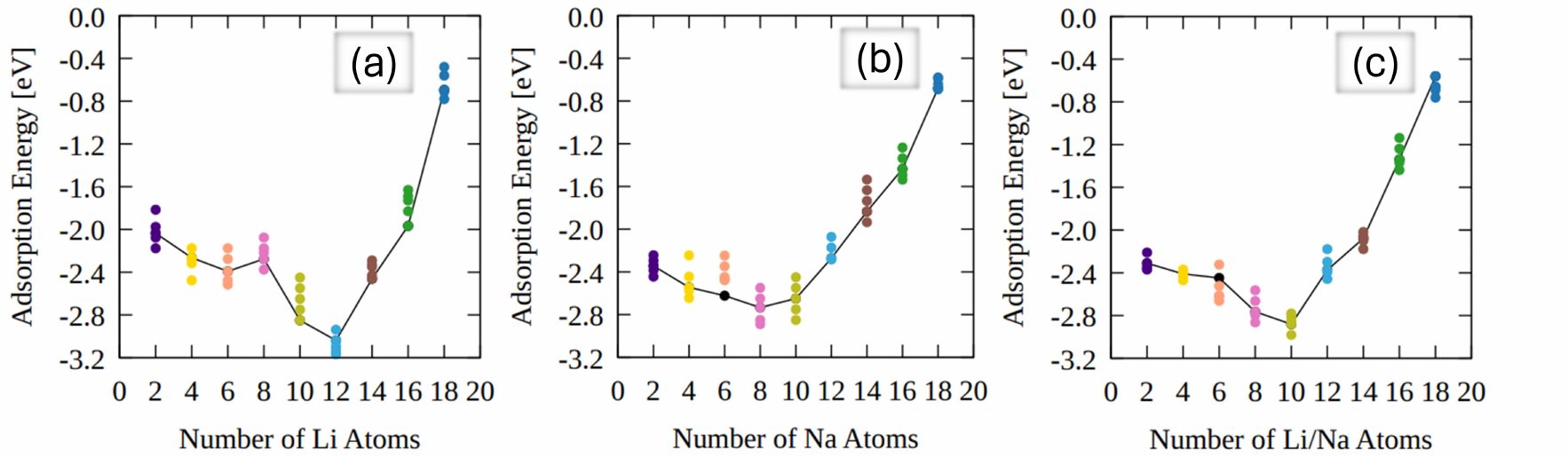}
    \caption{Evolution of adsorption energy as a function of the number of adsorbed ions on PCL monolayer. (a) The adsorption energy profile for \ce{Li} atoms, (b) \ce{Na} atoms, and (c) mixed \ce{Li}/\ce{Na} adsorption behavior.}
    \label{fig:adsorption}
\end{figure*}

Fig.~\ref{fig:adsorption} illustrates the evolution of the adsorption energy ($E_\text{ads}$) as a function of the number of \ce{Li}, \ce{Na}, and mixed \ce{Li}/\ce{Na} atoms adsorbed on the PLG surface. This analysis provides important insights into the binding strength, stability, and maximum storage capacity of ions within the material, which are essential for its performance as a battery anode.

The adsorption energy remains negative in all cases, confirming that \ce{Li} and \ce{Na} adsorption is thermodynamically favorable across all studied concentrations, suggesting favorable electrochemical performance. This trend has also been observed in other graphyne-based lattices \cite{yang2020nitrogen, singh2022alpha, nasrollahpour2018ab}. The adsorption energy curve exhibits a characteristic non-monotonic trend, where $E_\text{ads}$ initially decreases as more atoms are added, reaching a minimum value and gradually increasing as saturation limit is reached.

For \ce{Li} adsorption (Fig.~\ref{fig:adsorption}(a)), the most energetically favorable configuration occurs at approximately $12$ atoms, corresponding to the lowest value. Beyond this point, additional \ce{Li} atoms experience weaker binding due to reduced available adsorption sites and repulsive interactions, leading to a progressive increase in adsorption energy. A similar trend is observed for \ce{Na} adsorption (Fig.~\ref{fig:adsorption}(b)), though \ce{Na} exhibits slightly weaker binding than \ce{Li}, as reflected in its higher minimum adsorption energy. This difference aligns with the larger ionic radius of \ce{Na}, which affects its interaction with the PLG lattice.

The behavior of mixed \ce{Li}/\ce{Na} adsorption (Fig.~\ref{fig:adsorption}(c)) closely follows the trends observed for individual species, with a balanced interaction between \ce{Li} and \ce{Na} atoms. The co-adsorption mechanism suggests that PLG can simultaneously accommodate \ce{Li} and \ce{Na} without significant clustering or structural instability, making it a promising candidate for dual-ion battery applications. 

\begin{figure*}[pos=!htb]
    \centering
    \includegraphics[width=\linewidth]{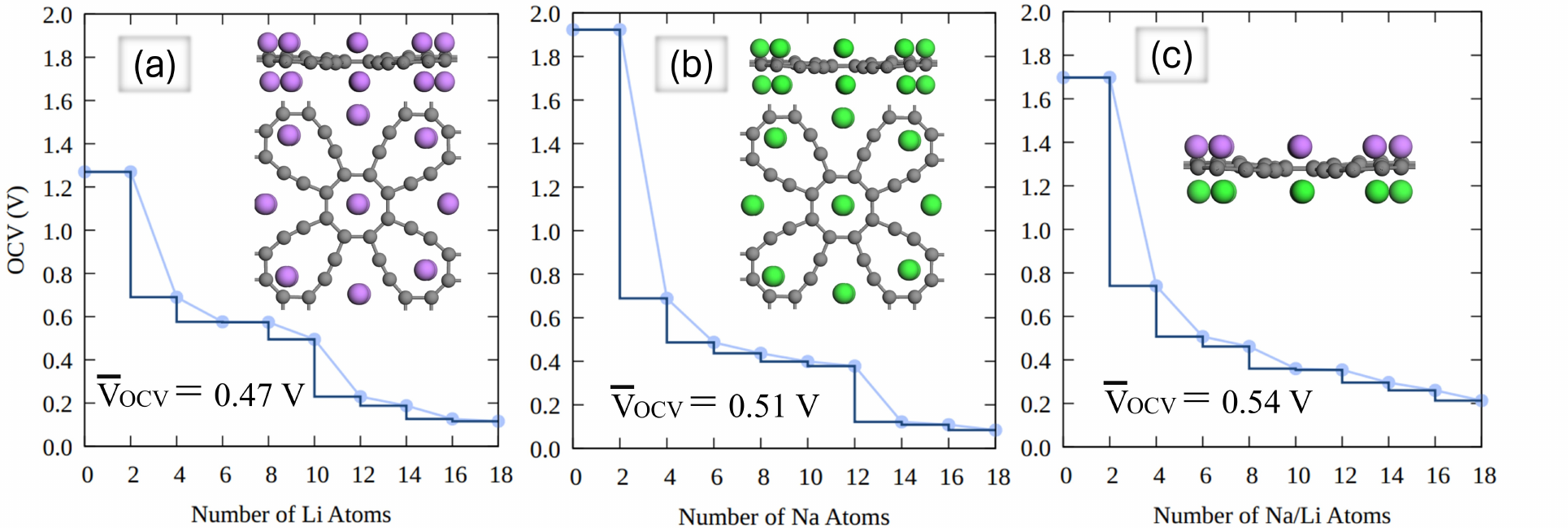}
    \caption{Open-circuit voltage (OCV) and \ce{Li} and \ce{Na} adsorption behavior in PLG. (a) OCV profile as a function of the number of adsorbed \ce{Li} atoms, with an average OCV of \SI{0.47}{\volt}. (b) OCV profile for \ce{Na} adsorption, showing an average OCV of \SI{0.51}{\volt}. (c) OCV profile for the mixed \ce{Li}/\ce{Na} system, with an average OCV of \SI{0.54}{\volt}. Insets illustrate the fully adsorbed configurations, where \ce{Li} (purple) and \ce{Na} (green) atoms occupy positions above and below the PCL monolayer.}
    \label{fig:ocv}
\end{figure*}

Another essential parameter for efficient battery performance is the OCV profile (see Fig.\ref{fig:ocv}). Here, the OCV values were obtained using the expression:

\begin{equation}
\text{OCV} = \frac{E_{\text{X-Graphyne}} + N E_{\text{M}} - E_{N-\text{M+X-Graphyne}}}{N e}
\end{equation}

\noindent where \( E_{\text{X-Graphyne}} \) is the total energy of pristine PLG, 
\( E_{\text{M}} \) represents the energy of a single metal atom (\ce{Li} or \ce{Na}), and \( E_{N-\text{M+X-Graphyne}} \) corresponds to the total energy of PLG with \( N \) adsorbed metal atoms. The parameter \( e \) denotes the elementary charge of an electron.

Figs.~\ref{fig:ocv}(a), \ref{fig:ocv}(b), and \ref{fig:ocv}(c) show the OCV evolution as a function of the number of \ce{Li}, \ce{Na}, and \ce{Li}/\ce{Na} adsorbed atoms. The OCV follows a stepwise decrease, which is expected due to the gradual occupation of adsorption sites and the corresponding change in interaction energies. The insets depict the fully adsorbed structures, where \ce{Li} and \ce{Na} atoms occupy positions above and below the PLG layer. At maximum metal loading, the OCV values are approximately \SI{0.2}{\volt} for all the systems. As widely reported \cite{zhang2019monolayer}, the OCV should remain low and positive, typically below \SI{1.0}{\volt}. Conversely, negative OCV values could lead to metal plating and result in battery failure. Regarding these features, PLG meets these requirements, ensuring stable and safe electrode performance.

\begin{figure*}[pos=!htb]
    \centering
    \includegraphics[width=\linewidth]{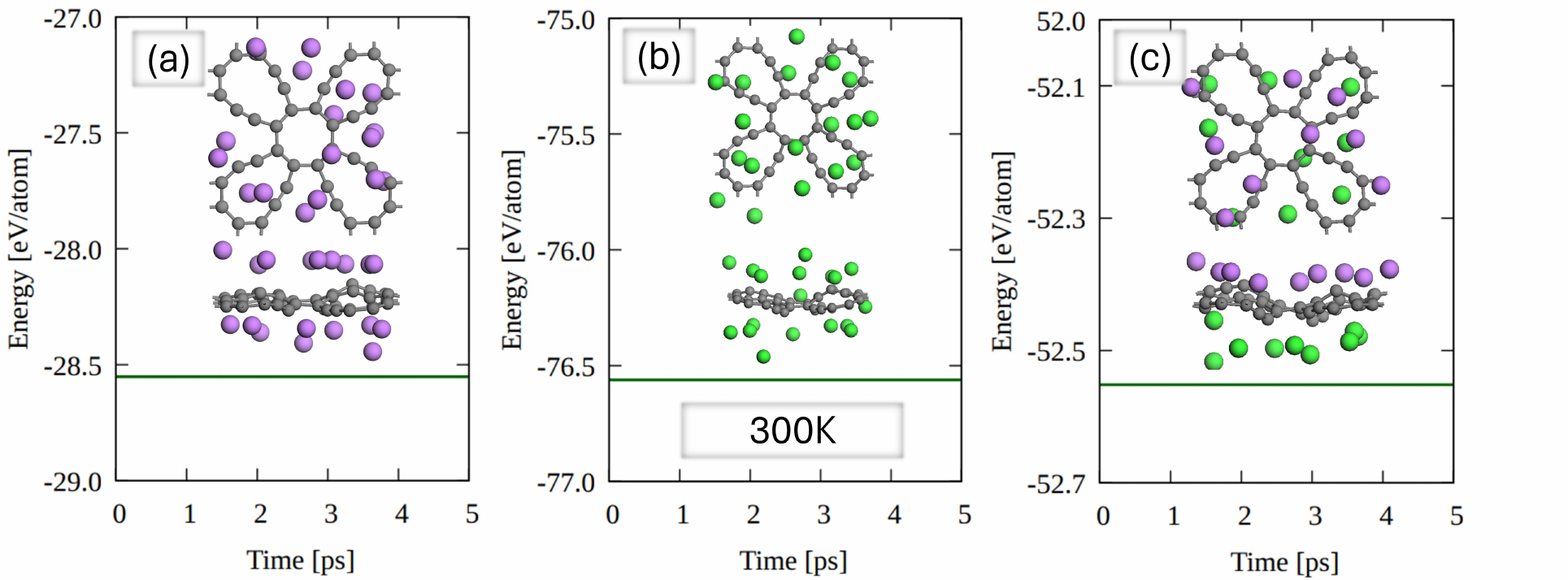}
    \caption{AIMD snapshots at \SI{300}{\kelvin} for (a) fully lithiated, (b) sodiated, and (c) mixed \ce{Li}/\ce{Na}-PLG systems.}
    \label{fig:aimdfull}
\end{figure*}

The thermal stability of fully lithiated and sodiated PLG is presented in Fig.~\ref{fid:aimdfull}. In this figure, the total energy per atom is monitored as a function of the simulation time, with the insets showing the final atomic configurations at the end of the AIMD simulations. In Fig.~\ref{fig:aimdfull}(a), the fully lithiated PLG system exhibits minimal energy fluctuations, with the total energy remaining nearly constant throughout the simulation. The final configuration confirms that \ce{Li} atoms remain well-distributed over the surface without significant clustering and detachment, indicating structural stability and strong \ce{Li} adsorption at the tested concentration values.

Similarly, Fig.~\ref{fig:aimdfull}(b) presents the AIMD results for the fully sodiated configuration. The system also maintains energetic stability over time, suggesting that \ce{Na} adsorption does not induce substantial lattice distortions. However, compared to \ce{Li} adsorption, Na exhibits slightly higher atomic displacements, likely due to its larger ionic radius and weaker binding energy, as previously observed in the adsorption energy analyses. Despite these minor displacements, no ion detachment was observed throughout the simulation, confirming the robustness of \ce{Na} adsorption in PLG. A similar trend is observed for the case considering the mixed \ce{Li}/\ce{Na} system (Fig.~\ref{fig:aimdfull}(c)). This configuration retains energetic stability comparable to the pure \ce{Li} and \ce{Na} cases, with no significant atomic reorganization or clustering. Importantly, all adsorbed ions remained anchored to the PLG surface, reinforcing the material’s capability to sustain dual-ion storage without ion desorption or structural instability.

Finally, the PLG theoretical storage capacity ($C$) was also evaluated by using the relation $C=N_{MAX}F/M$, where $N_{MAX}$ is the maximum number of adsorbed \ce{Li}/\ce{Na} atoms, $F$ is the Faraday constant (\SI{96485.332}{sA/mol}), and $M$ is the molar mass of the PLG supercell. Thus, the computed capacities are \SI{1004}{mAh/g} for \ce{Li}, \ce{Na}, and mixed \ce{Li}/\ce{Na} ions, surpassing traditional graphite (\SI{372}{mAh/g}) \cite{yi2024challenges,asenbauer2020success} and graphene anodes (\SI{744}{mAh/g}) \cite{yi2024challenges,gao2023twin}. Theoretical capacities for PLG, as presented here, are not only competitive but, in some cases, surpass those of other graphyne-related materials for LIBs/SIBs, such as $\gamma$-Graphyne (1037/570 mAh/g), \textit{N}-Graphdiyne ( 1660/623 mAh/g), penta-Graphyne (680/680 mAh/g), and Holey-Graphyne (873/558 mAh/g). In Table \ref{table:adsorption}, we provide a compilation of these results along with the corresponding energy barriers and data for other 2D carbon allotropes. These comparisons highlight the exceptional anodic performance of the PLG monolayer, particularly for \ce{Na}-ion batteries.

\begin{table*}[!ht]
\label{table:adsorption}
\centering
\caption{Comparison of the storage capacity ($C$) and in-plane energy barrier of PLG with other 2D materials used in LIBs and SIBs.} 
\begin{tabular*}{\linewidth}{@{\extracolsep{\fill}} lccc @{}}
\toprule
\textbf{System}                      & \textbf{$C$ (mAh/g)} & \textbf{Energy barrier (eV)} & \textbf{Reference} \\
\midrule
\textbf{Petal-Graphyne} & 1004/1004 & 0.28/0.25 & this work \\
\textbf{$\gamma$-Graphyne}  & 1037/570  & 0.44/0.46   &  \cite{yang2020nitrogen}  \\ 
\textbf{\textit{N}-Graphdiyne}  & 1660/623 & 0.80/0.79 & \cite{makaremi2018theoretical} \\
\textbf{Penta-Graphyne}  & 680/680 & 0.49/0.61 &  \cite{deb2022two} \\
\textbf{Holey-Graphyne} & 873/558 & 0.28/0.32 & \cite{sajjad2023two}   \\
\textbf{Biphenylene}  & 1302/1075  & 0.23/0.20 & \cite{duhan20232,han2022biphenylene}   \\
\textbf{Irida-graphene}   & 1116/1022  & 0.19/0.09  & \cite{xiong2024theoretical, martins2024irida}   \\ 
\textbf{T-graphene}  & 744/2232 & 0.37/0.35  & \cite{hu2020theoretical}  \\
\textbf{C$_5$N$_2$} & 914/914 & 0.24/0.17  & \cite{you20242d}  \\
\bottomrule
\end{tabular*}
\end{table*}

\section{Conclusions}

In summary, the structural, electronic, mechanical, optical, and energy storage properties of Petal-Graphyne (PLG), a newly proposed 2D carbon allotrope, were investigated. This novel structure exhibits metallic behavior, with multiple band crossings at the Fermi level, ensuring high electrical conductivity for electrochemical applications. The phonon dispersion and AIMD simulations confirm its thermal and dynamical stability, while its anisotropic mechanical properties reveal a maximum and minimum Young's modulus of \SI{180}{N/m} and \SI{60}{N/m}, respectively.

The optical analyses highlights strong absorption in the ultraviolet region, high reflectivity, and a refractive index profile consistent with metallic materials, making PLG a good candidate for optoelectronic applications. Additionally, the material demonstrates a high theoretical energy storage capacity of \SI{1004}{mAh/g} for both \ce{Li} and \ce{Na}-ions, outperforming conventional graphite, graphene, and other 2D carbon-based anodes. The average OCV values of \SI{0.47}{\volt} (\ce{Li}), \SI{0.51}{\volt} (\ce{Na}), and \SI{0.54}{\volt} (\ce{Li}/\ce{Na}) indicates stable electrochemical performance. The lowest in-plane diffusion barriers of \SI{0.28}{\electronvolt} (\ce{Li}) and \SI{0.25}{\electronvolt} (\ce{Na}) suggest good ion mobility, while the interlayer diffusion barriers of \SI{0.50}{\electronvolt} (\ce{Li}) and \SI{0.43}{\electronvolt} (\ce{Na}) confirm viable transport in stacked configurations.

\section*{Data access statement}
Data supporting the results can be accessed by contacting the corresponding author.

\section*{Conflicts of interest}
The authors declares no conflict of interest.

\section*{Acknowledgements}
This work was supported by the Brazilian funding agencies Fundação de Amparo à Pesquisa do Estado de São Paulo - FAPESP (grant no. 2022/03959-6, 2022/00349- 2, 2022/14576-0, 2020/01144-0, 2024/05087-1, and 2022/16509-9), and National Council for Scientific, Technological Development - CNPq (grant no. 307213/2021–8). L.A.R.J. acknowledges the financial support from FAP-DF grants 00193.00001808/2022-71 and $00193-00001857/2023-95$, FAPDF-PRONEM grant 00193.00001247/2021-20, PDPG-FAPDF-CAPES Centro-Oeste 00193-00000867/2024-94, and CNPq grants $350176/2022-1$ and $167745/2023-9$. A.C.D acknowledges the following FAP-DF grants: 00193-00001817/2023-43 and 00193-00002073/2023-84, the following CNPq grants: 408144/2022-0, 305174/2023-1, 444069/2024-0 and 444431/2024-1, and the computational resources from Cenapad-SP (project 897) and Lobo Carneiro HPC (project 133). K.A.L.L. and D.S.G. acknowledge the Center for Computational Engineering \& Sciences (CCES) at Unicamp for financial support through the FAPESP/CEPID Grant 2013/08293-7.

\printcredits

\bibliography{cas-refs}

\begin{thebibliography}{10}

\bibitem{tiwari2016magical}
Santosh~K Tiwari, Vijay Kumar, Andrzej Huczko, R~Oraon, A~De Adhikari, and GC~Nayak.
\newblock Magical allotropes of carbon: prospects and applications.
\newblock {\em Critical Reviews in Solid State and Materials Sciences}, 41(4):257--317, 2016.

\bibitem{nasir2018carbon}
Salisu Nasir, Mohd~Zobir Hussein, Zulkarnain Zainal, and Nor~Azah Yusof.
\newblock Carbon-based nanomaterials/allotropes: A glimpse of their synthesis, properties and some applications.
\newblock {\em Materials}, 11(2):295, 2018.

\bibitem{hirsch2010era}
Andreas Hirsch.
\newblock The era of carbon allotropes.
\newblock {\em Nature materials}, 9(11):868--871, 2010.

\bibitem{yi2018microporosity}
Wen-cai Yi, Wei Liu, Jorge Botana, Jing-yao Liu, and Mao-sheng Miao.
\newblock Microporosity as a new property control factor in graphene-like 2d allotropes.
\newblock {\em Journal of Materials Chemistry A}, 6(22):10348--10353, 2018.

\bibitem{zheng2015two}
Xiaoyu Zheng, Jiayan Luo, Wei Lv, Da-Wei Wang, and Quan-Hong Yang.
\newblock Two-dimensional porous carbon: synthesis and ion-transport properties.
\newblock {\em Advanced Materials}, 27(36):5388--5395, 2015.

\bibitem{paul2019carbon}
Rajib Paul, Mewin Vincent, Vinodkumar Etacheri, and Ajit~K Roy.
\newblock Carbon nanotubes, graphene, porous carbon, and hybrid carbon-based materials: synthesis, properties, and functionalization for efficient energy storage.
\newblock In {\em Carbon Based Nanomaterials for Advanced Thermal and Electrochemical Energy Storage and Conversion}, pages 1--24. Elsevier, 2019.

\bibitem{li2021rational}
Tongtong Li, Cheng He, and Wenxue Zhang.
\newblock Rational design of porous carbon allotropes as anchoring materials for lithium sulfur batteries.
\newblock {\em Journal of Energy Chemistry}, 52:121--129, 2021.

\bibitem{LIMA2025116099}
K.A.L. Lima, D.A. {da Silva}, G.D.~Amvame Nze, F.L.~Lopes de~Mendonça, M.L. Pereira, and L.A. Ribeiro.
\newblock Structural, electronic, and li-ion adsorption properties of polypygy explored by first-principles and machine learning simulations: A new multi-ringed 2d carbon allotrope.
\newblock {\em Journal of Energy Storage}, 117:116099, 2025.

\bibitem{you20242d}
Manqi You, Gencai Guo, Yujie Liao, Siwei Luo, Chaoyu He, Chao Tang, and Jianxin Zhong.
\newblock 2d novel c5n2 allotropes: High-performance anode materials for alkali metal ion battery.
\newblock {\em Journal of Energy Storage}, 84:111004, 2024.

\bibitem{xiong2024theoretical}
Xin Xiong, Hong-Bao Cao, Zheng Lu, Chun-Sheng Liu, and Xiao-Juan Ye.
\newblock Theoretical prediction on irida-graphene monolayer as promising anode material for lithium-ion batteries.
\newblock {\em Computational Materials Science}, 244:113225, 2024.

\bibitem{martins2024irida}
Nicolas~F Martins, Jos{\'e}~A Laranjeira, Guilherme~SL Fabris, Pablo~A Denis, and Julio~R Sambrano.
\newblock Irida-graphene as a high-performance anode for sodium batteries.
\newblock {\em Journal of Energy Storage}, 104:114637, 2024.

\bibitem{geim2007rise}
Andre~K Geim and Konstantin~S Novoselov.
\newblock The rise of graphene.
\newblock {\em Nature materials}, 6(3):183--191, 2007.

\bibitem{desyatkin2022scalable}
Victor~G Desyatkin, William~B Martin, Ali~E Aliev, Nathaniel~E Chapman, Alexandre~F Fonseca, Douglas~S Galv{\~a}o, Ericka~Roy Miller, Kevin~H Stone, Zhong Wang, Dante Zakhidov, et~al.
\newblock Scalable synthesis and characterization of multilayer $\gamma$-graphyne, new carbon crystals with a small direct band gap.
\newblock {\em Journal of the American Chemical Society}, 144(39):17999--18008, 2022.

\bibitem{fan2021biphenylene}
Qitang Fan, Linghao Yan, Matthias~W Tripp, Ond{\v{r}}ej Krej{\v{c}}{\'\i}, Stavrina Dimosthenous, Stefan~R Kachel, Mengyi Chen, Adam~S Foster, Ulrich Koert, Peter Liljeroth, et~al.
\newblock Biphenylene network: A nonbenzenoid carbon allotrope.
\newblock {\em Science}, 372(6544):852--856, 2021.

\bibitem{hou2022synthesis}
Lingxiang Hou, Xueping Cui, Bo~Guan, Shaozhi Wang, Ruian Li, Yunqi Liu, Daoben Zhu, and Jian Zheng.
\newblock Synthesis of a monolayer fullerene network.
\newblock {\em Nature}, 606(7914):507--510, 2022.

\bibitem{hudspeth2010electronic}
Mathew~A Hudspeth, Brandon~W Whitman, Veronica Barone, and Juan~E Peralta.
\newblock Electronic properties of the biphenylene sheet and its one-dimensional derivatives.
\newblock {\em ACS nano}, 4(8):4565--4570, 2010.

\bibitem{berber2004rigid}
Savas Berber, Eiji Osawa, and David Tom{\'a}nek.
\newblock Rigid crystalline phases of polymerized fullerenes.
\newblock {\em Physical Review B—Condensed Matter and Materials Physics}, 70(8):085417, 2004.

\bibitem{baughman1987structure}
RH~Baughman, H~Eckhardt, and M~Kertesz.
\newblock Structure-property predictions for new planar forms of carbon: Layered phases containing sp 2 and sp atoms.
\newblock {\em The Journal of chemical physics}, 87(11):6687--6699, 1987.

\bibitem{wallace1947band}
Philip~Richard Wallace.
\newblock The band theory of graphite.
\newblock {\em Physical review}, 71(9):622, 1947.

\bibitem{zhang2015penta}
Shunhong Zhang, Jian Zhou, Qian Wang, Xiaoshuang Chen, Yoshiyuki Kawazoe, and Puru Jena.
\newblock Penta-graphene: A new carbon allotrope.
\newblock {\em Proceedings of the National Academy of Sciences}, 112(8):2372--2377, 2015.

\bibitem{wang2018popgraphene}
Shuaiwei Wang, Baocheng Yang, Houyang Chen, and Eli Ruckenstein.
\newblock Popgraphene: a new 2d planar carbon allotrope composed of 5--8--5 carbon rings for high-performance lithium-ion battery anodes from bottom-up programming.
\newblock {\em Journal of Materials Chemistry A}, 6(16):6815--6821, 2018.

\bibitem{wang2015phagraphene}
Zhenhai Wang, Xiang-Feng Zhou, Xiaoming Zhang, Qiang Zhu, Huafeng Dong, Mingwen Zhao, and Artem~R Oganov.
\newblock Phagraphene: a low-energy graphene allotrope composed of 5--6--7 carbon rings with distorted dirac cones.
\newblock {\em Nano letters}, 15(9):6182--6186, 2015.

\bibitem{junior2023irida}
ML~Pereira J{\'u}nior, Wiliam~Ferreira da~Cunha, William~Ferreira Giozza, Rafael~Timoteo de~Sousa~Junior, and LA~Ribeiro Junior.
\newblock Irida-graphene: A new 2d carbon allotrope.
\newblock {\em FlatChem}, 37:100469, 2023.

\bibitem{tromer2023mechanical}
Raphael~M Tromer, Marcelo~L Pereira~Junior, Kleuton~A L.~Lima, Alexandre~F Fonseca, Luciano~R da~Silva, Douglas~S Galv{\~a}o, and Luiz~A Ribeiro~Junior.
\newblock Mechanical, electronic, and optical properties of 8-16-4 graphyne: A 2d carbon allotrope with dirac cones.
\newblock {\em The Journal of Physical Chemistry C}, 127(25):12226--12234, 2023.

\bibitem{mao2014nanocarbon}
Xianwen Mao, Gregory~C Rutledge, and T~Alan Hatton.
\newblock Nanocarbon-based electrochemical systems for sensing, electrocatalysis, and energy storage.
\newblock {\em Nano Today}, 9(4):405--432, 2014.

\bibitem{chen2023biphenylene}
Xin-Wei Chen, Zheng-Zhe Lin, and Xi-Mei Li.
\newblock Biphenylene network as sodium ion battery anode material.
\newblock {\em Physical Chemistry Chemical Physics}, 25(5):4340--4348, 2023.

\bibitem{ferguson2017biphenylene}
David Ferguson, Debra~J Searles, and Marlies Hankel.
\newblock Biphenylene and phagraphene as lithium ion battery anode materials.
\newblock {\em ACS applied materials \& interfaces}, 9(24):20577--20584, 2017.

\bibitem{santos2024proposing}
EAJ Santos, KAL Lima, and LA~Ribeiro~Junior.
\newblock Proposing todd-graphene as a novel porous 2d carbon allotrope designed for superior lithium-ion battery efficiency.
\newblock {\em Scientific Reports}, 14(1):6202, 2024.

\bibitem{ullah2024theoretical}
Saif Ullah, Marcos~G Menezes, and Alexander~M Silva.
\newblock Theoretical characterization of tolanene: A new 2d sp-sp2 hybridized carbon allotrope.
\newblock {\em Carbon}, 217:118618, 2024.

\bibitem{gomez2024tpdh}
Juan Gomez~Quispe, Bruno Ipaves, Douglas~Soares Galvao, and Pedro Alves da~Silva Autreto.
\newblock Tpdh-graphene as a new anodic material for lithium ion battery: Dft-based investigations.
\newblock {\em ACS omega}, 9(37):39195--39201, 2024.

\bibitem{liu2024novel}
Huili Liu, Yaru Wei, Donghai Wu, and Shuaiwei Wang.
\newblock A novel 2d carbon allotrope for high-performance metal-ion battery anode material.
\newblock {\em Materials Science in Semiconductor Processing}, 173:108146, 2024.

\bibitem{li2023thfs}
Tian-Kai Li, Xiao-Juan Ye, Lan Meng, and Chun-Sheng Liu.
\newblock Thfs-carbon: a theoretical prediction of metallic carbon allotrope with half-auxeticity, planar tetracoordinate carbon, and potential application as anode for sodium-ion batteries.
\newblock {\em Physical Chemistry Chemical Physics}, 25(22):15295--15301, 2023.

\bibitem{cheng2021two}
Zishuang Cheng, Xiaoming Zhang, Hui Zhang, Jianbo Gao, Heyan Liu, Xiao Yu, Xuefang Dai, Guodong Liu, and Guifeng Chen.
\newblock Two-dimensional metallic carbon allotrope with multiple rings for ion batteries.
\newblock {\em Physical Chemistry Chemical Physics}, 23(34):18770--18776, 2021.

\bibitem{li2017psi}
Xiaoyin Li, Qian Wang, and Puru Jena.
\newblock $\psi$-graphene: a new metallic allotrope of planar carbon with potential applications as anode materials for lithium-ion batteries.
\newblock {\em The journal of physical chemistry letters}, 8(14):3234--3241, 2017.

\bibitem{sun2024surface}
Luye Sun, Wei Zheng, Faming Kang, Wenze Gao, Tongde Wang, Guohua Gao, and Wei Xu.
\newblock On-surface synthesis and characterization of anti-aromatic cyclo [12] carbon and cyclo [20] carbon.
\newblock {\em Nature Communications}, 15(1):7649, 2024.

\bibitem{clark2005first}
Stewart~J Clark, Matthew~D Segall, Chris~J Pickard, Phil~J Hasnip, Matt~IJ Probert, Keith Refson, and Mike~C Payne.
\newblock First principles methods using castep.
\newblock {\em Zeitschrift f{\"u}r kristallographie-crystalline materials}, 220(5-6):567--570, 2005.

\bibitem{perdew1996generalized}
John~P Perdew, Kieron Burke, and Matthias Ernzerhof.
\newblock Generalized gradient approximation made simple.
\newblock {\em Physical review letters}, 77:3865, 1996.

\bibitem{grimme2006semiempirical}
Stefan Grimme.
\newblock Semiempirical gga-type density functional constructed with a long-range dispersion correction.
\newblock {\em Journal of computational chemistry}, 27(15):1787--1799, 2006.

\bibitem{baroni2001phonons}
Stefano Baroni, Stefano De~Gironcoli, Andrea Dal~Corso, and Paolo Giannozzi.
\newblock Phonons and related crystal properties from density-functional perturbation theory.
\newblock {\em Reviews of modern Physics}, 73(2):515, 2001.

\bibitem{nose1984unified}
Shuichi Nos{\'e}.
\newblock A unified formulation of the constant temperature molecular dynamics methods.
\newblock {\em The Journal of chemical physics}, 81:511--519, 1984.

\bibitem{metropolis1953equation}
Nicholas Metropolis, Arianna~W Rosenbluth, Marshall~N Rosenbluth, Augusta~H Teller, and Edward Teller.
\newblock Equation of state calculations by fast computing machines.
\newblock {\em The journal of chemical physics}, 21(6):1087--1092, 1953.

\bibitem{kirkpatrick1983optimization}
Scott Kirkpatrick, C~Daniel Gelatt~Jr, and Mario~P Vecchi.
\newblock Optimization by simulated annealing.
\newblock {\em science}, 220(4598):671--680, 1983.

\bibitem{vcerny1985thermodynamical}
Vladim{\'\i}r {\v{C}}ern{\`y}.
\newblock Thermodynamical approach to the traveling salesman problem: An efficient simulation algorithm.
\newblock {\em Journal of optimization theory and applications}, 45:41--51, 1985.

\bibitem{makri2019preconditioning}
Stela Makri, Christoph Ortner, and James~R Kermode.
\newblock A preconditioning scheme for minimum energy path finding methods.
\newblock {\em The Journal of Chemical Physics}, 150(9):094109, 2019.

\bibitem{barzilai1988two}
Jonathan Barzilai and Jonathan~M Borwein.
\newblock Two-point step size gradient methods.
\newblock {\em IMA journal of numerical analysis}, 8(1):141--148, 1988.

\bibitem{bitzek2006structural}
Erik Bitzek, Pekka Koskinen, Franz G{\"a}hler, Michael Moseler, and Peter Gumbsch.
\newblock Structural relaxation made simple.
\newblock {\em Physical review letters}, 97(17):170201, 2006.

\bibitem{PhysRevB.90.224104}
F\'elix Mouhat and Fran\ifmmode \mbox{\c{c}}\else \c{c}\fi{}ois-Xavier Coudert.
\newblock Necessary and sufficient elastic stability conditions in various crystal systems.
\newblock {\em Phys. Rev. B}, 90:224104, Dec 2014.

\bibitem{doi:10.1021/acs.jpcc.9b09593}
Yiran Ying, Ke~Fan, Sicong Zhu, Xin Luo, and Haitao Huang.
\newblock Theoretical investigation of monolayer rhtecl semiconductors as photocatalysts for water splitting.
\newblock {\em The Journal of Physical Chemistry C}, 124(1):639--646, 2020.

\bibitem{shao2012temperature}
Tianjiao Shao, Bin Wen, Roderick Melnik, Shan Yao, Yoshiyuki Kawazoe, and Yongjun Tian.
\newblock Temperature dependent elastic constants and ultimate strength of graphene and graphyne.
\newblock {\em The Journal of chemical physics}, 137(19), 2012.

\bibitem{pei2012mechanical}
Yang Pei.
\newblock Mechanical properties of graphdiyne sheet.
\newblock {\em Physica B: Condensed Matter}, 407(22):4436--4439, 2012.

\bibitem{aliev2025planar}
Ali~E Aliev, Yongzhe Guo, Alexandre~F Fonseca, Joselito~M Razal, Zhong Wang, Douglas~S Galv{\~a}o, Claire~M Bolding, Nathaniel~E Chapman-Wilson, Victor~G Desyatkin, Johannes~E Leisen, et~al.
\newblock A planar-sheet nongraphitic zero-bandgap sp2 carbon phase made by the low-temperature reaction of $\gamma$-graphyne.
\newblock {\em Proceedings of the National Academy of Sciences}, 122(5):e2413194122, 2025.

\bibitem{koo2014widely}
Jahyun Koo, Minwoo Park, Seunghyun Hwang, Bing Huang, Byungryul Jang, Yongkyung Kwon, and Hoonkyung Lee.
\newblock Widely tunable band gaps of graphdiyne: an ab initio study.
\newblock {\em Physical Chemistry Chemical Physics}, 16(19):8935--8939, 2014.

\bibitem{long2011electronic}
Mengqiu Long, Ling Tang, Dong Wang, Yuliang Li, and Zhigang Shuai.
\newblock Electronic structure and carrier mobility in graphdiyne sheet and nanoribbons: theoretical predictions.
\newblock {\em ACS nano}, 5(4):2593--2600, 2011.

\bibitem{li2020structural}
Linwei Li, Weiye Qiao, Hongcun Bai, and Yuanhe Huang.
\newblock Structural and electronic properties of $\alpha$-, $\beta$-, $\gamma$-, and 6, 6, 18-graphdiyne sheets and nanotubes.
\newblock {\em RSC advances}, 10(28):16709--16717, 2020.

\bibitem{li2023artificial}
Jiaqiang Li and Yu~Han.
\newblock Artificial carbon allotrope $\gamma$-graphyne: Synthesis, properties, and applications.
\newblock {\em Giant}, 13:100140, 2023.

\bibitem{sha2021first}
Xianwei Sha and Clifford~M Krowne.
\newblock First principles quantum calculations for graphyne for electronic devices.
\newblock {\em Nanoscale Advances}, 3(20):5853--5859, 2021.

\bibitem{muz2020electronic}
{\.I}skender Muz and Mustafa Kurban.
\newblock The electronic structure, transport and structural properties of nitrogen-decorated graphdiyne nanomaterials.
\newblock {\em Journal of Alloys and Compounds}, 842:155983, 2020.

\bibitem{hou2018study}
Xun Hou, Zhongjing Xie, Chunmei Li, Guannan Li, and Zhiqian Chen.
\newblock Study of electronic structure, thermal conductivity, elastic and optical properties of $\alpha$, $\beta$, $\gamma$-graphyne.
\newblock {\em Materials}, 11(2):188, 2018.

\bibitem{ge2018review}
Chuannan Ge, Jie Chen, Shaolong Tang, Youwei Du, and Nujiang Tang.
\newblock Review of the electronic, optical, and magnetic properties of graphdiyne: from theories to experiments.
\newblock {\em ACS applied materials \& interfaces}, 11(3):2707--2716, 2018.

\bibitem{jafarzadeh2020electronic}
Hamed Jafarzadeh, Saeed Zahedi, and Amir~Hossein Bayani.
\newblock Electronic and optical properties of 14, 14, 18 graphyne as an anti-visible ray coating.
\newblock {\em Optik}, 203:163905, 2020.

\bibitem{duhan20232}
Nidhi Duhan, Brahmananda Chakraborty, and TJ~Dhilip Kumar.
\newblock 2-dimensional biphenylene monolayer as anode in li ion secondary battery with high storage capacity: Acumen from density functional theory.
\newblock {\em Applied Surface Science}, 629:157171, 2023.

\bibitem{han2022biphenylene}
Ting Han, Yu~Liu, Xiaodong Lv, and Fengyu Li.
\newblock Biphenylene monolayer: a novel nonbenzenoid carbon allotrope with potential application as an anode material for high-performance sodium-ion batteries.
\newblock {\em Physical Chemistry Chemical Physics}, 24(18):10712--10716, 2022.

\bibitem{hu2020theoretical}
Junping Hu, Yu~Liu, Ning Liu, Jianwen Li, and Chuying Ouyang.
\newblock Theoretical prediction of t-graphene as a promising alkali-ion battery anode offering ultrahigh capacity.
\newblock {\em Physical Chemistry Chemical Physics}, 22(6):3281--3289, 2020.

\bibitem{gao2023twin}
Shuli Gao, Elyas Abduryim, Changcheng Chen, Chao Dong, Xiaoning Guan, Shuangna Guo, Yue Kuai, Ge~Wu, Wen Chen, and Pengfei Lu.
\newblock Twin-graphene: a promising anode material for lithium-ion batteries with ultrahigh specific capacity.
\newblock {\em The Journal of Physical Chemistry C}, 127:14065--14074, 2023.

\bibitem{zhang2013graphdiyne}
Hongyu Zhang, Yueyuan Xia, Hongxia Bu, Xiaopeng Wang, Meng Zhang, Youhua Luo, and Mingwen Zhao.
\newblock Graphdiyne: A promising anode material for lithium ion batteries with high capacity and rate capability.
\newblock {\em Journal of Applied Physics}, 113(4), 2013.

\bibitem{bartolomei2024sodium}
Massimiliano Bartolomei and Giacomo Giorgi.
\newblock Sodium into $\gamma$-graphyne multilayers: An intercalation compound for anodes in metal-ion batteries.
\newblock {\em ACS materials letters}, 6(10):4682--4689, 2024.

\bibitem{lemaalem2023graphyne}
Mohammed Lemaalem, Nabil Khossossi, Gaelle Bouder, Poulumi Dey, and Philippe Carbonni{\`e}re.
\newblock Graphyne-based membrane as a promising candidate for li-battery electrodes protection: Insight from atomistic simulations.
\newblock {\em Journal of Power Sources}, 581:233482, 2023.

\bibitem{lee2022understanding}
Hyobin Lee, Seungwon Yang, Suhwan Kim, Jihun Song, Joonam Park, Chil-Hoon Doh, Yoon-Cheol Ha, Tae-Soon Kwon, and Yong~Min Lee.
\newblock Understanding the effects of diffusion coefficient and exchange current density on the electrochemical model of lithium-ion batteries.
\newblock {\em Current Opinion in Electrochemistry}, 34:100986, 2022.

\bibitem{li2014graphdiyne}
Yongjun Li, Liang Xu, Huibiao Liu, and Yuliang Li.
\newblock Graphdiyne and graphyne: from theoretical predictions to practical construction.
\newblock {\em Chemical Society Reviews}, 43(8):2572--2586, 2014.

\bibitem{yang2020nitrogen}
Chaofan Yang, Chong Qiao, Yang Chen, Xueqi Zhao, Lulu Wu, Yong Li, Yu~Jia, Songyou Wang, and Xiaoli Cui.
\newblock Nitrogen doped $\gamma$-graphyne: A novel anode for high-capacity rechargeable alkali-ion batteries.
\newblock {\em Small}, 16(10):1907365, 2020.

\bibitem{singh2022alpha}
Tavinder Singh, Jyoti~Roy Choudhuri, and Malay~Kumar Rana.
\newblock $\alpha$-graphyne as a promising anode material for na-ion batteries: a first-principles study.
\newblock {\em Nanotechnology}, 34(4):045404, 2022.

\bibitem{nasrollahpour2018ab}
Mokhtar Nasrollahpour, Mohsen Vafaee, Mohammad~Reza Hosseini, and Hossein Iravani.
\newblock Ab initio study of sodium diffusion and adsorption on boron-doped graphyne as promising anode material in sodium-ion batteries.
\newblock {\em Physical Chemistry Chemical Physics}, 20(47):29889--29895, 2018.

\bibitem{zhang2019monolayer}
Xiuying Zhang, Chen Yang, Yuanyuan Pan, Mouyi Weng, Linqiang Xu, Shiqi Liu, Jie Yang, Jiahuan Yan, Jingzhen Li, Bowen Shi, et~al.
\newblock Monolayer gas with high ion mobility and capacity as a promising anode battery material.
\newblock {\em Journal of Materials Chemistry A}, 7(23):14042--14050, 2019.

\bibitem{yi2024challenges}
Xiaoping Yi, Guoqing Qi, Xunliang Liu, Christopher Depcik, and Lin Liu.
\newblock Challenges and strategies toward anode materials with different lithium storage mechanisms for rechargeable lithium batteries.
\newblock {\em Journal of Energy Storage}, 95:112480, 2024.

\bibitem{asenbauer2020success}
Jakob Asenbauer, Tobias Eisenmann, Matthias Kuenzel, Arefeh Kazzazi, Zhen Chen, and Dominic Bresser.
\newblock The success story of graphite as a lithium-ion anode material--fundamentals, remaining challenges, and recent developments including silicon (oxide) composites.
\newblock {\em Sustainable Energy \& Fuels}, 4(11):5387--5416, 2020.

\bibitem{makaremi2018theoretical}
Meysam Makaremi, Bohayra Mortazavi, Timon Rabczuk, Geoffrey~A Ozin, and Chandra~Veer Singh.
\newblock Theoretical investigation: 2d n-graphdiyne nanosheets as promising anode materials for li/na rechargeable storage devices.
\newblock {\em ACS Applied Nano Materials}, 2(1):127--135, 2018.

\bibitem{deb2022two}
Jyotirmoy Deb, Rajeev Ahuja, and Utpal Sarkar.
\newblock Two-dimensional pentagraphyne as a high-performance anode material for li/na-ion rechargeable batteries.
\newblock {\em ACS Applied Nano Materials}, 5(8):10572--10582, 2022.

\bibitem{sajjad2023two}
Muhammad Sajjad, Khaled Badawy, J~Andreas Larsson, Rehan Umer, and Nirpendra Singh.
\newblock Two dimensional holey graphyne: An excellent anode and anchoring material for metal--ion and metal--sulfur batteries.
\newblock {\em Carbon}, 214:118340, 2023.

\end{thebibliography}

\end{document}